\documentclass[11pt]{article}  

\usepackage{amssymb,amstext,amsmath,amsthm}
\usepackage[dvips]{graphicx}
\usepackage{latexsym}
\usepackage{psfrag}
\usepackage{amsfonts}
\usepackage{bbm}
\usepackage{color}

\newcommand{\Ref}[1]{(\ref{#1})}

\newcommand{\eqa}{\begin{eqnarray}}
\newcommand{\neqa}{\end{eqnarray}}
\newcommand{\equ}{\begin{equation}}
\newcommand{\nequ}{\end{equation}}

\def\om{\omega}
\def\w{\wedge}

\def\ra{\rangle}

\newcommand{\ket}[1]{|{#1}\ra}

\newcommand{\p}{\partial}

\def\d{\delta}
\def\f{\frac}
\def\tl{\tilde}
\def\wtl{\widetilde}

\usepackage{bbm}
\def\C{{\mathbbm C}}

\newcommand{\scr}{\rm\scriptscriptstyle}
\newcommand{\su}{\mathit{su}}
\newcommand{\so}{\mathit{so}}

\def\vp{\varphi}
\let\eps=\epsilon
\def\th{\theta}

\newcommand{\pb}{\Phi_{\beta}}
\newcommand{\der}[2]{\f{\d{#1}}{\d{#2}}}
\newcommand{\mb}{\bar{m}}
\newcommand{\rnm}{r_{\scr NM}}
\newcommand{\rH}{r_{\scr H}}
\newcommand{\GN}{G_{\scr N}}

\begin{document}

\title{\Large\bf Spherically symmetric black holes \\ 
in minimally modified self-dual gravity}
\author{{Akihiro Ishibashi${}^{ab}$ and Simone Speziale${}^{bc}$}\footnote{akihiro.ishibashi@kek.jp, simone.speziale@cpt.univ-mrs.fr}
\\
{\small ${}^a$\emph{Institute of Particle and Nuclear Studies, KEK, Ibaraki 305-0801, Japan}}\\
{\small ${}^b$\emph{Perimeter Institute for Theoretical Physics, 31 Caroline St. N, ON N2L 2Y5, Waterloo,
Canada}} \\ %{\small \emph{and}} \\
{\small ${}^c$\emph{Centre de Physique Th\'eorique de Luminy, Case 907, 13288, Marseille, France}}}
\date{\today}

\maketitle

\begin{abstract}
We discuss spherically symmetric black holes in the modified self-dual theory 
of gravity recently studied by Krasnov, obtained adding a Weyl-curvature 
dependent `cosmological term' to the Plebanski lagrangian for general 
relativity. 
This type of modified gravity admits two different types of singularities: 
one is a true singularity for the theory where the fundamental fields of 
the theory, as well as the (auxiliary) spacetime metric, become singular, 
and the other one is a milder ``non-metric singularity'' where 
the metric description of the spacetime breaks down but the fundamental 
fields themselves are regular. 
We first generalise this modified self-dual gravity to include Maxwell's 
field and then study basic features of spherically symmetric, charged black 
holes, with particular focus on whether these two 
types of singularities are hidden or naked. 
We restrict our attention to minimal forms of the modification, and 
find that the theory exhibits `screening' effects of the electric charge 
(or `anti-screening', depending upon the sign of the modification term), 
in the sense that it leads to the possibility of 
charging the black hole more (or less) than it would be possible in general 
relativity without exposing a naked singularity.  
We also find that for any (even arbitrarily large) value of charge, 
true singularities of the theory appear to be either achronal (non-timelike) 
covered by the hypersurface of a harmless non-metric singularity, 
or simply hidden inside at least one Killing horizon. 

\end{abstract}

\tableofcontents

%%%%%%%%%%%
\section{Introduction}
%%%%%%%%%%%n

Ever since general relativity was formulated, various different types of 
modifications of the theory have been proposed, aiming at, e.g., 
obtaining better control of gravity in quantum regime or attempting to account 
for astrophysical/cosmological observations without invoking unknown, 
exotic matter fields in the cosmic inventory.   
% 
% \medskip 
Recently, Krasnov has introduced yet another modification of general 
relativity \cite{Krasnov,Krasnov1,Krasnov2,Krasnov3,Krasnov4,Krasnov5} 
where the standard cosmological constant term is made into a function $\Phi$ of the Weyl curvature, 
in such a fashion as not to add extra degrees of freedom. The absence of extra degrees of freedom is a remarkable 
property which distinguishes the case at hand from usual modifications of gravity.
The key to this result is to give up the metric as the fundamental field of 
the theory (hence the definition ``non-metric theory of gravity'' used for 
instance in \cite{Krasnov,Krasnov1,Krasnov2}), and to use the formulation of 
general relativity in terms of a self-dual 2-form as the fundamental field. 
This idea was introduced long ago by Plebanski \cite{Plebanski}, and later developed by 
Capovilla, Dell, Jacobson and Mason \cite{Capo, Capo2}.
A metric can be derived from this 2-form through the imposition 
of suitable constraints. Krasnov's main result is that these constraints can 
be generalised to a class of theories labelled by the choice of $\Phi$, 
all of which give dynamics for a metric with only two degrees of freedom. 
Such generalisation had already been considered in the Hamiltonian framework 
by Bengtsson and Peldan \cite{Bengtsson,Bengtsson-Peldan}. 
See also \cite{Capo3,Smolin,nuovi}. 

\medskip 
It is tempting to speculate that such a modification of general relativity 
gives an effective description of quantum gravity models based on the 
Plebanski lagrangian, e.g. the spin foam formalism for loop quantum 
gravity~\cite{Carlo}. In this paper, however, we consider the theory only 
from a classical viewpoint, with the goal of investigating the physical 
consequences of the modification, and its viability as a theory of 
classical gravity. 

\medskip 
The nature of the modification makes gravity behave differently in regions of 
different Weyl curvature. 
As the Weyl curvature vanishes on conformally flat metrics, such as Minkowski or FLRW 
cosmological spacetimes, these are also exact solutions in the modified theory. The effect of the modification can be studied perturbing around these solutions, but one needs to go to second order in perturbation theory. 
It is then easier to investigate the classical effects of the modification for
spherical symmetric solutions with non-zero Weyl curvature, where departures 
from general relativity are expected right away. 
This case has been considered by Krasnov and Shtanov \cite{Krasnov2}. 
They looked at specific choices of $\Phi$ without matter, and 
found solutions analogue of the Schwarzschild black hole.\footnote{A central motivation of Krasnov and Shtanov \cite{Krasnov2,Krasnov5} is to investigate whether the modification to gravity can be suitable to 
explaining the anomalous rotational curve of spiral galaxies without 
appealing to dark matter, and they have considered profiles of $\Phi$ tailored to that end.}  
A characteristic feature discovered in \cite{Krasnov2} is the existence of a new class of singularities,
where the derived metric is singular but the fundamental fields of the theory remain finite.

\medskip 
In this paper we extend the analysis of the theory to the coupling 
to electromagnetism. In spite of the derived nature of the metric field, 
a natural coupling in terms of the fundamental 2-form was found by Capovilla, 
Dell, Jacobson and Mason \cite{Capo2}, and as anticipated by Krasnov 
\cite{Krasnov1}, such coupling can be extended straightforwardly to the 
modified theory. This allows us to study spherically symmetric electro-vacuum 
black hole solutions, for which we prove staticity for general form of $\Phi$. 
We then restrict our attention to the two simplest profiles of $\Phi$, and 
give explicit solutions. We describe how the asymptotic structure of 
the solutions depends on $\Phi$, and for one of the two chosen profiles 
we find the standard Reissner-Nordst\o m-de Sitter asymptotic structure. 
On such a solution, we investigate its geometry near the horizon and the way 
this is affected by the modification.

\medskip
We find that the non-metric singularities are still present and their location 
depends on the value of the charge. The number and locations of the horizons 
are also affected by the modification. With respect to general relativity, 
we find that more/less charge can be poured into the hole before the event 
horizon disappears, depending upon the sign of the modifying term: 
more charge when the sign of the modification is coherent with a negative 
cosmological constant, and less charge when it is coherent with a positive 
cosmological constant. In a sense, one could say that the electric 
charge is screened in one case, and anti-screened in the other case. 
In particular, in the first case, a sector of the theory exists where 
arbitrarily charged solutions admit a (Killing) horizon which is located 
inside a non-metric singularity (for this reason, it cannot be identified with the event horizon) and covers the central singularity at $r=0$. 
This striking departure from general relativity shows that useful bounds
on the physically allowed types of $\Phi$ can be obtained by 
looking at classical solutions coupled with matter. 

\medskip 
The plan of the paper is as follows. 
In order to keep the paper as self-contained as possible, 
and also to establish our notation and conventions, the next two Sections 
are devoted to reviews of self-dual gravity theories; 
Pleba\'{n}ski's self-dual formulation of general relativity in the next Section 
and Krasnov's modified self-dual gravity in Sect.~\ref{SecModPleb}.   
We also refer the reader to the literature provided in the bibliography 
for more details on both the standard Pleba\'{n}ski formulation of gravity 
and the modification proposed by Krasnov. 
Section \ref{SecMax} describes the coupling to electromagnetism. 
In Section \ref{SecSph} we restrict attention to the spherical symmetric case, 
and show how to obtain the field equations. 
We roughly follow \cite{Krasnov2}, although the notation and 
presentation differ here and there. 
We also reprove Birkhoff's theorem in a slightly more conventional way. 
In Section \ref{SecPhi}, we discuss the rationale for $\Phi$ and possible 
links to the existence of asymptotically flat solutions. 
In Section \ref{sec: MG: Quadratic} we discuss the field equations for 
the minimal modification which allows asymptotic flatness, 
and the existence of Reissner-Nordstr\o m and Nariai-Bertotti-Robinson 
solutions. We then focus on this minimal modification; 
in Section \ref{SecVacuum} we review the analytic solution for the vacuum 
black hole found by Krasnov and Shtanov in \cite{Krasnov2}, and 
in Section \ref{SecRN} we present the solutions for the charged case. 
In the final Section \ref{SecConc} we give an overview of the properties of 
the solutions. Throughout the paper, we set $c=4\pi\eps_0=1$.

%%%%%%%%%%%
\section{Self-duality and the Pleba\'{n}ski Action}\label{sect:Plebanski}
%%%%%%%%%%%
The notion of self-duality plays a key role in the Pleba\'{n}ski construction of 
general relativity. 
Consider a 2-form $F=\f12 \, F_{\mu\nu} \, dx^\mu \w dx^\nu$ on a pseudo-Riemannian manifold with metric
$g_{\mu\nu}$ and non-zero determinant $g<0$. Introduce 
the identity ${\mathbbm 1} \equiv \d^{\mu\nu}_{\rho\sigma}\equiv \d^\mu_{[\rho}\d^{\nu}_{\sigma]}$
(here and in the following a square bracket on the indices means normalised anti-symmetrisation) and
the Hodge star
$\star \equiv \f1{2\sqrt{-g}}\, \eps^{\mu\nu}{}_{\rho\sigma}$
(here $\eps$ is the completely antisymmetric tensor density with $\eps^{0123}=1$). The latter
satisfies $\star^2=-1$, as can be seen immediately using
\equ\label{epseps} 
\eps^{\rho\sigma}{}_{\mu\nu} \, \eps^{\mu\nu}{}_{\lambda\tau} 
= 4 \, g \, \d^{\rho\sigma}_{\lambda\tau} \,.
\nequ
Then, the complex orthogonal projectors ${\cal P}^{\pm} = \f12 ({\mathbbm 1} \pm i \star)$
(i.e. ${\cal P}^{\pm}{}^2 = {\cal P}^{\pm}$, ${\cal P}^{\pm}{\cal P}^{\mp}=0$)
define respectively the self-dual and antiself-dual parts of the 2-form $F$,
\equ\label{defSD}
F^{(\pm)}_{\mu\nu} = \big({\cal P}^{\pm}\, F\big)_{\mu\nu} 
 = \f12 \bigg( F_{\mu\nu} \pm \f{i}{2\sqrt{-g}} \, \eps^{\rho\sigma}{}_{\mu\nu} \, F_{\rho\sigma} \bigg) \,.
\nequ
These are respectively the positive and negative eigenvectors of $i \star$,
\equ
i \star F^{(\pm)} = \pm F^{(\pm)} \,, 
\nequ
and we have $F = F^{(+)} + F^{(-)}.$\footnote{We put the label $\pm$ of self and antiself-dual form within
a bracket, to avoid confusion with SU(2) indices later on. In the literature one also finds the convention
${\cal P}^{\pm} = \f12 ({\mathbbm 1} \mp i \star)$, so that self and antiself are the $\pm i$ eigenvectors 
of $\star$.} An important property which will play a role in the modified theory
is the invariance of $\star$ under conformal transformations of the metric $g_{\mu\nu}$.

\medskip
In the following, we will work with forms valued in the Lorentz algebra $\so(3,1)$. A typical example is
the tetrad $e^I_\mu(x)$, defined by $g_{\mu\nu} = e^I_\mu e^J_\nu \eta_{IJ}$. The tetrad is
a 1-form with indices $I=0 \ldots 3$ in the fundamental representation of the Lorentz algebra. 
Another well-known example is the spin connection $\om_\mu^{IJ}(e)$, a 1-form with values in the adjoint representation of $\so(3,1)$. For an object in the adjoint representation, we can define the algebraic self-dual and antiself-dual
components, as we did in \Ref{defSD}. This time we have to use the Hodge operator on the Lorentz bundle, which reads $\f12\eps^{IJ}{}_{KL}$; and thus
${\cal P}^{\pm} = \f12 (\d^{IJ}_{KL} \pm \f{i}2 \eps^{IJ}{}_{KL})$.
These two projectors realize explicitly the familiar homomorphism $\so(3,1;\C)=\su_{\scr R}(2)\oplus\su_{\scr L}(2)$
of the Lorentz algebra into two $\su(2)$ algebras, which rather than self-dual and antiself-dual 
are more commonly dubbed right-handed and left-handed.
To make the mapping more explicit, it is convenient to pick out the time direction $I=0$, and define
$\om^{(\pm)}{}^i \equiv \om^{(\pm)}{}^{0i}$, with $i=1,2,3$ an SU(2) index.
(Anti)self-duality then means $\om^{(\pm)}{}^{0i}= \pm \f{i}2 \, \eps^i{}_{jk} \, \om^{(\pm)}{}^{jk}$.
 
\medskip
Consider now a 2-form with values in the adjoint representation, $B^{IJ}_{\mu\nu}$.
For this object we can consider the self-dual projections in both spacetime and algebra indices.
Pleba\'{n}ski \cite{Plebanski, Capo2} found that the two notions of self-duality coincide if the 2-form is constrained to satisfy the following quadratic condition,
\equ\label{metrConstr}
B^i \w B^j -\f13 \d^{ij} B^k \w B_k = 0 \,.
\nequ
This equation is called metricity constraint, and it is solved by $B^i\propto \Sigma^i(e)$, where
\equ\label{Sigma}
\Sigma^i(e) =  e^0 \w e^i + \f{i}{2}\, \eps^i{}_{jk} \,e^j \w e^k 
\nequ
is (twice) the self-dual part of the two form $e\w e$. It satisfies\footnote{Note that also the antiself-dual part
of $e\w e$ solves \Ref{metrConstr}. The two solutions are distinguished by the sign of the right hand side of \Ref{e}.}
\equ\label{e}
\Sigma^i \w \Sigma^j =2\, i \, e \, \d^{ij} \, d^4x \,,
\nequ
with $e=\eps^{\mu\nu\rho\sigma} e^0_\mu e^1_\nu e^2_\rho e_\sigma^3$ 
the determinant of the tetrad, related to the determinant of the metric by det $g_{\mu\nu}=-e^2$.
The proportionality coefficient between $B$ and $\Sigma$ can thus be fixed to 1 requiring
\equ\label{det}
B^i \w B_i = 6\, i \, e \, d^4x \,.
\nequ

\medskip
Therefore a $B$ field solving \Ref{metrConstr} encodes a metric via the tetrad field in \Ref{Sigma}. 
In order for these to give a real Lorentzian metric, reality conditions on $B$
have also to be imposed. These are \cite{Capo2} 
\equ\label{real}
B^i \w {B}^*{}^j = 0, \qquad {\rm Im} \, (i B^i \w {B}_i)=0,
\nequ
which imply respectively that $B^*{}^i$ is proportional to the antiself-dual form, 
and that the determinant \Ref{det} is real. 
The $B$ field itself can well be complex.
These additional reality conditions are the trade-off for using the simpler self-dual variables
instead of the full Lorentz group. 

\medskip 
In the following, we will refer to a $B$ field satisfying \Ref{metrConstr}, \Ref{det} and \Ref{real}
and the reality conditions as ``metric''.

\medskip
To come to the dynamics, let us also introduce an SU(2) connection $\om^i_\mu$ 
-- as an independent field, not the spin connection -- and its curvature 2-form 
$F^i_{\mu\nu}(\om) = d \om^i + \f{i}{2} \eps^i{}_{jk} \om^j \w \om^k$.
Including the constraint \Ref{metrConstr} with a symmetric and traceless Lagrange multiplier $\Psi_{ij}$, 
the action
\equ\label{actionPleb}
S_{\rm G}[B, \om, \Psi] = \f1{8\pi \GN} 
\int \left[B^i \w F_i(\om) + \f12 \Big(\Psi_{ij} - \f13 \Lambda \d_{ij} \Big) B^i \w B^j\right]
\nequ
is equivalent, once the reality conditions for the fields are imposed, to ($i$ times) the Einstein-Hilbert action for general relativity with cosmological constant $\Lambda$, 
plus a total divergence \cite{Plebanski, Capo}. It is instructive to see the
equivalence at the level of the equations of motion.
Next to \Ref{metrConstr}, the other field equations are
\eqa\label{EqCart}
&& d B^i + i \eps^i{}_{jk} \om^j \w B^k = 0 \,, \\
\label{EqF1}
&& F^i(\om) + \Big(\Psi^i{}_j-\f13\Lambda \d^i{}_j \Big) B^j = 0 \,.
\neqa
When $B$ is metric, \Ref{EqCart} becomes the first Cartan structure equation, which 
-- upon requiring invertibility of $e_\mu^I$ -- identifies $\om$ as the (self-dual part of the) spin connection $\om(e)$. This in turns implies that $F^i(\om(e))$ is the (self-dual part of the) Riemann tensor (via the second Cartan
structure equation). 
Finally, the 18 equations \Ref{EqF1} have the following double role: ten of them 
give the Einstein equations for the tetrad, while five identify the Lagrange multiplier $\Psi^{ij}$ with the 
(self-dual) Weyl part of the Riemann tensor. The remaining three equations are gauge.
See \cite{Krasnov1, Capo2, KrasnovPleb} and references therein for more details.

\medskip 
Before moving on, let us add a brief comment on Pleba\'{n}ski's theory.
The metric solution \Ref{Sigma} to the constraints is unique if $B^i \w B_i \neq 0$ is required, a condition
that translates into the invertibility of the tetrad. 
Degenerate solutions with $B^i \w B_i =0$ also exist, and they have no analogue in general relativity.
The existence of this other sector makes Pleba\'{n}ski's gravity more general than Einstein's theory.
Let us also recall that the canonical analysis of \Ref{actionPleb} gives the Ashtekar formalism \cite{Ashtekar}, which is 
at the roots of loop quantum gravity \cite{Carlo}.

%%%%%%%%%%%
\section{Modified Pleba\'{n}ski gravity and non-metricity}\label{SecModPleb} 
%%%%%%%%%%%

The modification introduced by Krasnov is to promote the cosmological 
constant term in \Ref{actionPleb} 
to a scalar function of the Lagrange multiplier itself, say  $\Phi(\Psi)$.
The action then reads
\equ\label{extP}
S[B, \om, \Psi] = \f1{8\pi \GN} 
\int \left[B^i \w F_i(\om) + \f12\Big(\Psi_{ij}+ \d_{ij} \, \Phi(\Psi)\Big) B^i \w B^j\right] \,.
\nequ
As before, $i=1,2,3$ is an SU(2) index, $\Psi_{ij}$ is symmetric and traceless, and 
the fields are complex. The reality conditions are the same as before.
For the moment, we leave $\Phi$ arbitrary.

\medskip
When $\Phi = \Phi_0$ is constant we obtain Pleba\'{n}ski's gravity with $\Lambda=-3 \Phi_0$;
a non-constant $\Phi$ is responsible for deviations from general relativity.
To understand the nature of the deviations, let us look at the field equations. The $\om$ and $B$ variations
yield
\eqa\label{EqCart2}
&& d B^i + i \eps^i{}_{jk} \om^j \w B^k = 0 \,, \\
\label{EqF2}
&& F^i(\om) + \Big(\Psi^i{}_j + \Phi(\Psi) \d^i{}_j \Big) B^j = 0 \,. 
\neqa
The key difference with Pleba\'{n}ski's gravity comes from the variation with respect to the field $\Psi$.
In order to take this variation, notice that only two algebra scalars can be built out of a symmetric and traceless $\Psi_{ij}$, $z_1 \equiv {\rm Tr}\, \Psi^2 = \Psi^i{}_{j} \Psi^j{}_i$ 
and $z_2 \equiv {\rm Tr}\, \Psi^3 = \Psi^i{}_j \Psi^j{}_{k} \Psi^k{}_{i}$. Therefore $\Phi$ is truly
a function of $z_1$ and $z_2$. Taking also into account the tracelessness of $\Psi$, the variation gives
\equ\label{dP}
\left(\der{\Phi(\Psi)}{\Psi} \right)^{ij} =
2 \, \Psi^{ij} \, \der{\Phi}{z_1} + 3 \Big(\Psi^i{}_m \, \Psi^{mj} 
- \f13 \, \d^{ij} \, {\rm Tr}\, \Psi^2 \Big) 
\der{\Phi}{z_2} \,.
\nequ
Thus instead of the metricity constraints \Ref{metrConstr}, we now get the following equation,
\equ\label{EqB}
B^i \w B^j - \f13 \d^{ij} B^k \w B_k \, = \, - \left(\der{\Phi(\Psi)}{\Psi}\right)^{ij} \, B^k \w B_k \,.
\nequ

\medskip
When the right hand side is non-zero, \Ref{Sigma} is no longer a solution, and the theory is inequivalent to
general relativity. Hence the quantity $\d \Phi/\d \Psi$ controls the departures
from general relativity.
In the previous Section we recalled that when the right hand side of \Ref{EqB} vanishes, 
the solution $B^i=\Sigma^i(e)$ encodes a metric $e_\mu^I$ through \Ref{Sigma}. 
We are now going to show how a metric is encoded in $B^i$ also when the right
hand side does not vanish.

%--------------------------------------
\subsection{Extracting the metric from the modified constraints}\label{SecSolving}
%--------------------------------------

The equations \Ref{EqB} are the key to understand the nature of the modification to gravity. 
They are manifestly not anymore constraints for $B$, but rather five equations\footnote{The
trace of \Ref{EqB} vanishes identically.} relating $B$ and $\Psi$. 
For instance, they can be used to determine the five $\Psi^{ij}$'s as functions of the $B^i$'s, 
once a specific choice of $\Phi$ is made.  
The $\Psi$ so obtained can be then plugged in the equations \Ref{EqF2} to obtain the dynamics of $B$. 

\medskip
There is however an alternative procedure, as emphasised by Krasnov \cite{Krasnov1}, that allows us to
describe the theory in more familiar metric terms. 
This is based on the fact that the notion of Hodge self-duality naturally defines a metric
(up to a conformal factor), even before imposing \Ref{metrConstr}. Hence,
simply declaring $B$ to be self-dual suffices to endow the theory with a metric, and one can
derive its field equations from the action \Ref{extP}.

\medskip
To be more explicit, consider the following two symmetric tensors, $g_{\mu\nu}^{\scr U}$ and $h^{ij}$,
that can be constructed out of $B^i$ \cite{KrasnovNuovo,nuovi, Urbantke},
\eqa \label{gU}
\sqrt{|g^{\scr U}|} \, g_{\mu\nu}^{\scr U} &=& \f1{12} \,\eps^{\alpha\beta\gamma\d} \, \eps_{ijk} \, 
B^i_{\mu\alpha} B^j_{\beta\gamma} B^k_{\d\nu} \,, 
\\ \label{h}
B^i \w B^j &=& \f13 B^k \w B_k \ h^{ij} \,, \qquad h^k{}_k = 3\,.
\neqa
Here $g^{\scr U}$ is the determinant of the Urbantke metric $g^{\scr U}_{\mu\nu}$ introduced in \cite{Urbantke,Capo2}.
The crucial fact \cite{KrasnovNuovo,nuovi} is that whenever $h^{ij}$ is invertible, 
$B^i$ is self-dual with respect to the Urbantke metric \Ref{gU} -- or to any metric related to it
by a conformal transformation, because of the invariance of $\star$.
%Invertibility of $h^{ij}$ is in turn guaranteed by the non-degeneracy of $B^i\w B^j$, i.e. $B^i\w B_i\neq 0$.
Hence the notion of self-duality together with $B^i\w B_i\neq 0$ suffices to have a unique metric,
up to conformal transformations; we do not need to impose any constraints.
%The above result still holds for $V\neq 1$, the only difference being that 
%the metric with respect to which $B^i$ is self-dual is not exactly \Ref{gU}, 
%but it is still related to it simply by a conformal transformation. 

\medskip
But if a metric is already present, or rather a conformal class of metrics, what is the 
role of the metricity constraints \Ref{metrConstr} in the standard theory? It is instructive at this point
to count the variables. A generic triple of 2-forms $B^i$ has $3\times 6=18$ components,
of which $3$ can be gauged away using the SU(2) symmetry. The remaining 15, thanks to the above result and
when $B^i\w B^j$ is non-degenerate, can be parametrised in terms of the two tensors
\Ref{gU} and \Ref{h}: $B^i=B^i(g_{\mu\nu}^{\scr U}, h^{ij})$.
Then, the role of the five equations \Ref{EqB} is to fix the five components of $h^{ij}$: to constants in the standard
case, to functions of $\Psi^{ij}$ in the modified case.

\medskip
To see how the fixing goes, consider first the standard case. Plugging \Ref{h} into the metricity constraints \Ref{metrConstr}, we get $h^{ij}=\d^{ij}$.
This kills the extra five components in the $(g^{\scr U}_{\mu\nu}, h^{ij})$ parametrisation of $B^i$, and we are left 
with only the metric $g_{\mu\nu}^{\scr U}$ as the variable of the theory. 
An explicit computation then shows that $g_{\mu\nu}^{\scr U}$ coincides with the metric $e_\mu^I e_\nu^J \eta_{IJ}$
encoded in $\Sigma^i(e)$ via the Pleba\'{n}ski solution \Ref{Sigma}.
The only conformal ambiguity left is the rescaling by a constant, but this can be fixed as in \Ref{det},
or reabsorbed in the definition of $\GN$ and $\Lambda$.
So the standard constraints guarantee the existence of a preferred metric in the conformal class.

\medskip
In the modified case when the right hand side of \Ref{EqB} is non-vanishing, we can still use these equations
to eliminate $h^{ij}$, but rather than being fixed proportional to the identity, this time its components
satisfy algebraic relations with (the derivatives of) the $\Psi^{ij}$'s. Thus after solving \Ref{EqB} we have a
new parametrisation $B^i(g^{\scr U}_{\mu\nu}, \Psi^{ij})$. Plugging the latter 
in the torsionless condition \Ref{EqCart2} and solving this equation\footnote{That this is
possible in general is shown e.g. in \cite{Deser,Halpern,Bengtsson2,KrasnovNuovo,nuovi}.} 
we obtain a ($\Psi$-dependent) spin connection $\om(g^{\scr U}, \Psi^{ij})$. Finally, inserting the solutions
$B(g^{\scr U}, \Psi^{ij})$ and $\om(g^{\scr U}, \Psi^{ij})$ into 
\Ref{EqF2} gives us the dynamics for $g^{\scr U}_{\mu\nu}$ and $\Psi^{ij}$.

\medskip
However, the conformal ambiguity is not trivial anymore in the modified case, for 
we can also consider metrics which differ from \Ref{gU} by a $\Psi$-dependent conformal factor.
One such metric arises if we solve the modified metricity constraints in terms of the metric encoded in $\Sigma(e)$,
instead of \Ref{gU}. 
As remarked in \cite{Krasnov1}, this alternative construction is more convenient for practical reasons,
and it is the one that we use in this paper.
%
%\medskip
%As remarked in \cite{Krasnov1}, it is indeed possible and more convenient to solve the modified
%metricity constraints in terms of the metric encoded in $\Sigma(e)$. 
%This alternative construction, which is the one that we use in this paper, goes as follows.
First of all, we restrict to the $B^i\w B_i\neq 0$ sector, and assume $h^{ij}$
to be invertible, and thus diagonalizable since symmetric.
Consider the matrix $D^i{}_j$ of normalised eigenvectors, which is such that 
$h^{i}{}_{j}=D^i{}_{k} \d^k{}_l (D^{-1})^{l}{}_j$.
$h^{ij}$ has five independent components, hence also the diagonalizing matrix $D$. 
Call them $c_n$, $n=1\ldots 5$. Using $D(c_n)$ we can bring \Ref{h} in a form
reminiscent of the metricity constraint \Ref{metrConstr}. Defining
$\hat B^i = (D^{-1})^i{}_j B^j$, 
equation \Ref{h} becomes
\equ
\hat B^i \w \hat B^j = \f13\,\hat B^k\w \hat B_k \, \d^{ij} \,.
\nequ
This equation is solved by $\hat B^i = \Sigma^i(e)$, and therefore
\equ\label{sol}
B^i = D^i{}_j(c_n) \Sigma^j(e) \,.
\nequ
%When we write the solution in this form, we shift the five $h^{ij}$ components of $B^i$ into
%the coefficients $c_n$, thus obtaining a parametrisation $B^i (e_\mu^I, c_{n})$.
Inserting this ansatz for $B^i$ into \Ref{EqB} gives us algebraic relations between $c_n$ and 
(the derivatives of) $\Psi^{ij}$, from which we obtain $D^i{}_j(c_n(\Psi))\equiv \tl D^i{}_j(\Psi)$.
This gives a parametrization $B^i=B^i (e_\mu^I, \Psi^{ij})$ which solves \Ref{EqB}.

\medskip
This way of constructing a solution to the equations \Ref{EqB} has the advantage of a simple link to 
the usual Pleba\'{n}ski form \Ref{Sigma}. 
Using this $B$ to evaluate the Urbantke metric \Ref{gU} one obtains 
$g_{\mu\nu}^{\scr U} = f(\Psi) \, e_\mu^I e_\nu^J \eta_{IJ}$, where $f$ is a scalar function obtained from contractions
of three matrices $\tl D^i{}_j(\Psi)$.
As anticipated above, the metric $e_\mu^I e_\nu^J \eta_{IJ}$ contained in \Ref{Sigma} and the Urbantke metric \Ref{gU} now differ by a non-trivial conformal factor which depends on $\Psi$.

\medskip
Which of these two should be taken as the physical metric?  
As it turns out, there is no physically distinguished metric in the modified theory, and
only the conformal class of the metric is specified.\footnote{For a discussion on this,
see the literature by Krasnov and in particular \cite{KrasnovSmall}.} 
Thus both $e_\mu^Ie_\nu^J \eta_{IJ}$ and $g_{\mu\nu}^{\scr U}$,
as well as any other metric in the same class, are treated on equal footing.
This conformal ambiguity distinguishes the modified theory from Pleba\'{n}ski's.

\medskip
The ambiguity is expected to be resolved by matter couplings, and
indeed in \cite{KrasnovSmall} a mechanism to distinguish a physical metric was shown,
based on the requirement that test particles follow geodesics.
In this paper on the other hand we consider only the vacuum theory, or its coupling to electromagnetism,
which being conformally invariant is insensitive to the ambiguity.
Thus the ambiguity is fully present, and it becomes
relevant when discussing the singularities of the theory, because a metric has to be 
singled out in order to do so. To that end, we will consider a specific metric in the Sections on black holes below. 
We anticipate here that this is the metric
$e_\mu^Ie_\nu^J \eta_{IJ}$ which arises naturally in our construction of
the spherically symmetric solution.

%%%%%%%%
\subsection{Reality conditions and degrees of freedom}
%%%%%%%%

One would expect the procedure just described to expose the presence of new propagating degrees of freedom with respect to general relativity. 
Remarkably, the Hamiltonian analysis performed in \cite{Krasnov4} (see also \cite{Bengtsson}) shows that this is not the
case: also the modified theory has only 2 propagating degrees of freedom.
The key for the absence of new physical degrees of freedom is in the fact that the field equations 
for $\Psi^{ij}$ obtained from \Ref{EqF2} are purely algebraic \cite{KrasnovNuovo,nuovi}.
The property is related to the peculiar complex nature of the action, and 
to the fact that reality conditions have to be additionally imposed, by hand. 
It is ultimately this imposition that it is responsible for the absence of extra degrees of freedom,
both in the Pleba\'{n}ski theory \Ref{actionPleb} and in the modified theory \Ref{extP}.

\medskip
In this perspective, let us add a remark on the reality constraints before continuing. 
After the revival of Pleba\'{n}ski's complex formalism by Capovilla, Dell, Jacobson and Mason, it was later realized by 
Reisenberger \cite{Mike} (see also \cite{De Pietri}) that one can use real fields. 
In order to do so, one gives up the self-duality of the fields
and takes $B^{IJ}$, $\om^{IJ}$ and $\Psi^{IJKL}$ in the full Lorentz group.
The action is identical to \Ref{actionPleb} with Lorentz indices, and it is often called 
SO(3,1) (or non-chiral) Pleba\'{n}ski action. The absence of reality constraints is a manifest advantage of this 
formulation.

\medskip
The same modification proposed by Krasnov in \Ref{extP} can be applied to the non-chiral SO(3,1) action.
In its simplest form $\Phi \propto {\rm Tr}\, \Psi^2$, this was introduced by Smolin \cite{Smolin}.
However, although the chiral and non-chiral Pleba\'{n}ski actions lead to the same classical physics, this
might not be true for the modified actions. Above we mentioned how crucial the reality conditions are to ensure
that the modified theory \Ref{extP} only has two propagating degrees of freedom. The non-chiral action on the
other hand does not have the reality conditions, thus the modification 
might include new degrees of freedom. This was recently pointed out by Alexandrov and Krasnov in \cite{Alexandrov},
where a counting of the constraints in the Hamiltonian framework showed the presence of six additional
degrees of freedom.
The modified non-chiral action thus appears to be a rather different theory, and its relevance for gravity still to be explored.

%%%%%%%%
\section{Maxwell action}\label{SecMax}
%%%%%%%%
The coupling of matter to the standard Pleba\'{n}ski action \Ref{actionPleb} was 
studied in \cite{Capo2}, where actions for scalar, spinor and vector fields 
were proposed starting from a reformulation of the conventional curved-space 
matter lagrangians in terms of the fundamental fields $B^i$ and $\om^i$. 
This conservative approach guarantees that when the $B$ field is metric, 
the standard minimal coupling of matter fields to the metric is recovered. 

\medskip 
Such a reformulation is particularly natural for gauge theories, whose action 
can be written directly in terms of the spacetime Hodge star 
$\star= (1/{2e}) \eps^{\mu\nu}{}_{\rho\sigma}$. 
Thanks to this fact, the gauge theory lagrangian proposed in \cite{Capo2} 
can be immediately coupled to the modified action \Ref{extP}, and we will 
restrict our attention to this case.
\footnote{ %%% 
The actions for scalar and fermion fields proposed in \cite{Capo2} can not be 
coupled to \Ref{extP} the way they are, because they carry additional 
constraints, which are compatible only with the standard Pleba\'{n}ski ones 
\Ref{metrConstr} and not with \Ref{EqB} \cite{Krasnov3}. 
Hence suitable modifications of these actions have to be studied before 
scalars or fermions can be coupled to the modified theory.
} %%% 
In units $4\pi \eps_0=1$, the Maxwell action reads
\equ\label{M}
S(g_{\mu\nu}, A_\mu) = \f1{8\pi} \int F \w \star F =
-\f1{16\pi} \int d^4x \sqrt{-g} \, g^{\mu\rho} \, g^{\nu\sigma}\, F_{\mu\nu}(A) \, F_{\rho\sigma}(A) \,.
\nequ
To express this action in terms of the $B$ field, a simple choice is\footnote{Alternatives for the coupling
of Yang-Mills theory to the non-chiral SO(3,1) Pleba\'{n}ski action
have been considered in \cite{Smolin,Speziale}.} \cite{Capo2}
\equ\label{Max}
S_{\rm M}[\vp, A, B] = \f{1}{2\pi} \int \Big[\vp_i \, B^i \w F(A) - \f12 \, \vp_i \, \vp_j \, B^i \w B^j\Big] \,.
\nequ
This is a first order formulation of electromagnetism, where the vector potential $A$ is flanked
by an auxiliary field $\vp$. This action, which can be straighforwardly generalized to the
Yang-Mills non-abelian case, is reminiscent of the first order formulation of Yang-Mills theory in
flat space, sometimes called field strength formulation, or BFYM theory \cite{Halpern,BFYM}.

\medskip 
When $B$ is metric, \Ref{Max} is equivalent to ($i$ times) the Maxwell action \Ref{M}. To see this,
note that the vanishing of the $\vp$-variation gives
\equ\label{dSdvp}
B^i \w \Big[F(A) - \vp_j B^j \Big] = 0 \,.
\nequ
Recall that self and antiself-dual forms make a basis in the space of 2-forms. 
When the realitity conditions \Ref{real} hold, a self-dual $B$ implies that its complex conjugate $B^*$ is antiself-dual.
Then, a generic Maxwell field can be decomposed as
\equ\label{FBB}
F = F^{(+)}+F^{(-)} = F^{(+)}_i B^i + F^{(-)}_i B^*{}^i,
\nequ
and thus $B \w F %= B \w (F^{(+)} + F^{(-)}) 
\equiv B \w F^{(+)}$. Hence \Ref{dSdvp} implies
\equ\label{eqphi}
F^{(+)}_{\rho\sigma}(A) = \vp_j B^j_{\rho\sigma} \,.
\nequ
Substituting this result back into \Ref{Max} we obtain
\eqa\label{Max1}
S_{\rm M}[\vp, A, B] &=& \f{1}{16\pi} \int d^4x \eps^{\mu\nu\rho\sigma} F^{(+)}_{\mu\nu}(A) F^{(+)}_{\mu\nu}(A)
=  \f1{8\pi} \int F \w (\mathbbm 1 -i\star) F \,.
\neqa

\medskip 
Finally, if $B^i$ is metric we can use  \Ref{epseps} and get explicitly
\equ\label{M3}
%\Big[ F_{\rho\sigma}(A)- \f12 F^+_{\rho\sigma}(A) \Big] \no 
S_{\rm M}[\vp, A, B] = \f{-i}{16\pi} \int d^4x 
\Big[ \sqrt{-g} \, g^{\mu\rho} g^{\nu\sigma} F_{\mu\nu}(A) F_{\rho\sigma}(A)
- \f{i}2 \, \eps^{\mu\nu\rho\sigma} F_{\mu\nu}(A) F_{\rho\sigma}(A)\Big] \,.
\nequ
As promised, the first term of \Ref{M3} is ($i$ times) the Maxwell lagrangian \Ref{M}. 
The second term is a total divergence, and as such does not
contribute to the equations of motion.\footnote{The second term is the 
standard CP-violating $\theta_{\rm QCD}$-term, but with an $i$ factor that makes it purely imaginary.
As explained in \cite{Capo2}, this complex action for Maxwell theory can be obtained from the usual
one in the Hamiltonian framework through a complex canonical transformation, where the momentum
conjugated to the connection is $E+i B$.
This is analogue to what happens in the gravitational sector with Ashtekar's self-dual variables, 
where the momentum conjugated to the densitized triad is $\Gamma + i K$ \cite{Ashtekar}.}
Neglecting this total divergence (and the similar one arising in the gravity sector), 
\Ref{actionPleb} plus \Ref{Max} are equivalent to
\equ
S_{\rm tot} = i \left[
\f1{16\pi G} \int d^4x \sqrt{-g} \, R - \f1{16\pi}\int d^4x \sqrt{-g} \, g^{\mu\rho} \, g^{\nu\sigma}\, F_{\mu\nu}\,F_{\rho\sigma} \right]
\nequ 
in agreement with the standard conventions \cite{Gravitation}.

\medskip 
This proves the validity of the action \Ref{Max} for coupling Maxwell theory 
to Pleba\'{n}ski's gravity. 
The constraints \Ref{dSdvp} are compatible with the metricity constraints \Ref{metrConstr} as well as with
the modified constraints \Ref{EqB}, thus the action \Ref{Max} can also be coupled to the modified theory.
This is what we do in this paper. That is, we consider \Ref{Max} plus \Ref{extP} as the total action for modified gravity
coupled to electromagnetism.

\medskip 
Since the Maxwell action \Ref{Max} does not depend on the spin connection $\om$ nor the field $\Psi$, the coupled action
leads to the field equations \Ref{EqCart2} and \Ref{EqB} unchanged.
%(fermionic fields on the other hand are a source of torsion and thus can be expected to enter \Ref{EqCart2}).
The only equations to acquire a source are \Ref{EqF2}, which now read
\equ\label{eqFsource}
F^i(\om) + \big(\Psi^{i}{}_{j}+\Phi(\Psi) \d^{i}{}_{j} \big) B^j = 8\pi \GN \, \tau^i
\nequ
where we conveniently defined 
\begin{equation}
\der{S_{\rm matter}}{B^i_{\mu\nu}} 
 = - \f14 \eps^{\mu\nu\rho\sigma} \tau^i_{\rho\sigma} \,.
\end{equation} 
The algebra-valued 2-form $\tau^i=\f12\tau^i_{\mu\nu} dx^\mu\w dx^\nu$ 
plays the role of the energy-momentum tensor in the standard metric formalism, and we shall refer to
it as such. Notice however that unlike the standard energy-momentum tensor, $\tau^i$ is 
\emph{antisymmetric} in the spacetime indices.
For $S_{\rm matter}=S_{\rm M}$ given by \Ref{Max} we get
\equ\label{MT}
\tau_{\rho\sigma}^i = - \f1{2\pi} \vp^i F^{(-)}_{\rho\sigma}(A) \,.
\nequ
This satisfies $B^i \w \tau_i = 0$, the analogue of the standard tracelessness condition.

%%%%%%%%%%%
\section{Spherically symmetric spacetime}\label{SecSph}
%%%%%%%%%%%
In the rest of the paper we consider the spherically symmetric reduced sector of the theory.
This sector has been investigated by Krasnov and Shtanov \cite{Krasnov2,Krasnov5}, who obtained the 
reduced field equations in vacuum.
In this Section we review their results and extend them to the coupling to electromagnetism. 

%%%%%%%%%%%
\subsection{Preliminaries: changing SU(2) basis}
%%%%%%%%%%%
Let us look back at the solution \Ref{Sigma} of the standard Pleba\'{n}ski theory, which encodes a metric 
$ds^2 = e^0 \otimes e^0 - e^1 \otimes e^1 - e^2 \otimes e^2 - e^3 \otimes e^3$
via the tetrad $e_\mu^I$.
With the prospect of spherical symmetry in mind, it is convenient to work with a Newman-Penrose tetrad
$\ell_\mu^I = (l_\mu, n_\mu, m_\mu, \mb_\mu)$, which defines the metric in the form
\equ\label{gNP}
ds^2  = l\otimes n + n \otimes l - m \otimes \bar{m} - \bar{m} \otimes m,
\qquad l \w n \w m \w \mb = - i \, e \, d^4x \,.
\nequ
The transformation between tetrad basis is $l = (e^0-e^1)/\sqrt{2}$, $n= (e^0+e^1)/\sqrt{2}$,
$m = (e^2+ ie^3)/\sqrt{2}$, and $\mb = m^*$.

\medskip 
With the Newman-Penrose tetrad, the Pleba\'{n}ski solution \Ref{Sigma} reads
\equ\label{SigmaSpin}
\Sigma^1 = l \w n - m \w \mb \,, \quad \Sigma^2 =   n \w \mb + l \w m \,, 
\quad  \Sigma^3 = i (n \w \mb - l \w m) \,.
\nequ
This triple can be simplified with a change of basis in the internal SU(2) space.
With respect to the standard basis $\tau_i$ given by (one half times) the Pauli matrices, we consider
the spherical basis $\tl\tau_0 = \tau_3$, $\tl\tau_\pm = \pm (\tau_1 \pm i \tau_2)$.
In this new basis we have, after a trivial permutation $123\mapsto 312$ of the algebra indices, 
\equ\label{tildeSigma}
\wtl \Sigma^0 = l \w n - m \w \mb \,, \quad \wtl \Sigma^+ = n \w \mb \,, 
\quad \wtl\Sigma^- = m \w l \,. 
\nequ
A straightforward computation checks that this triple is still self-dual with respect 
to \Ref{gNP}, and that the reality conditions \Ref{real} are satisfied provided $l$ and $n$ are real
and $\mb = m^*$. Just as \Ref{SigmaSpin}, the triple
\Ref{tildeSigma} provides a basis in the space of self-dual 2-forms. 

\medskip 
In the spherical basis $\tl\tau_i$, the Killing-Cartan metric is not diagonal any longer, instead
\equ\label{tildeTrace}
{\rm Tr} (\tl\tau_i \tl\tau_j) = \f12 \, \tl\d_{ij} \,, 
\qquad
\tl\d_{ij} = \left(\begin{array}{ccc} 1 & 0 & 0 \\ 0 & 0 & -2 \\ 0 & -2 & 0 \\  \end{array}\right) \,.
\nequ
Accordingly we have
\equ\label{TraceNew}
{\rm Tr} (\Sigma \w \Sigma) = \f12 \, \wtl\Sigma^0 \w \wtl\Sigma^0 - 2 \, \wtl\Sigma^+ \w \wtl\Sigma^- \,.
\nequ
Consistency with \Ref{e} can be straightforwardly verified,
\equ
\d_{ij} \, \Sigma^i \w \Sigma^j = \tl\d_{ij} \, \wtl\Sigma^i \w \wtl\Sigma^j  = 
-6 \, l \w n \w m \w \mb = 6 \, i \, e \, d^4x \,.
\nequ

\medskip 
The non-trivial scalar product given by $\tl\d_{ij}$ is the trade-off for 
the simpler form of \Ref{tildeSigma}. However the advantage of the spherical 
basis $\tl\tau_i$ is not only to simplify the triple of self-dual 2-forms from 
\Ref{SigmaSpin} to \Ref{tildeSigma}, but also to allow us to deal with tensor 
product of irreps using the familiar Clebsch-Gordan decomposition.  
This will be needed to describe $\Psi$ and the modified constraint equation 
\Ref{EqB}. Before doing so, notice that the $\tl\tau_i$ are a basis in the 
adjoint representation ${\bf 1}$, but they are not normalised with respect to 
the scalar product \Ref{tildeTrace}. The transformation to the standard 
normalised basis $\ket{j,m}$ is: 
$\ket{1,0}=\ket{\tl\tau_0}$, $\ket{1,\pm 1}=\ket{\tl\tau_\pm}/\sqrt{2}$.  
These extra factors of $\sqrt{2}$ have to be taken into account when writing 
the Clebsch-Gordan coefficients.\footnote{ %%% 
Alternatively, one can work with the normalised SU(2) 
basis. This, however, introduces factors of $\sqrt{2}$ in \Ref{tildeSigma}, 
complicating the field equations.
} %%% 

%----------------------------------------------------------------------
\subsection{Solving the modified metricity constraints}\label{sect:constraints}
%----------------------------------------------------------------------
In order to solve the equations \Ref{EqB}, we need to write explicitly the right hand side.
To do so, recall that $\Psi$, which is a symmetric and traceless tensor product of two SU(2) adjoints, 
lives in the five dimensional irreducible representation ${\bf 2}$. To find a basis in this space, we consider
the (reducible) tensor product of two adjoints,
${\bf 1}\otimes {\bf 1} = {\bf 2} \oplus {\bf 1} \oplus {\bf 0}$. Here the irrep ${\bf 2}$
is the symmetric and traceless part of the product, 
the ${\bf 1}$ is the antisymmetric part, and the singlet ${\bf 0}$ is 
the trace. 
A basis in the irrep ${\bf 2}$ can be obtained using the following Clebsch-Gordan decomposition, 
\begin{subequations}\begin{align}\label{T}
&\ket{2,\pm 2} = \ket{1,{\pm1}} \otimes \ket{1,{\pm1}}, \\
&\ket{2,\pm 1} = \f1{\sqrt{2}} \Big(\ket{1,\pm 1} \otimes \ket{1,0} + \ket{1,0} \otimes \ket{1,\pm 1}\Big), \\
&\ket{2,0} = \f1{\sqrt{6}} \Big(\ket{1,1}  \otimes \ket{1,-1} + \ket{1,-1} \otimes \ket{1,1}
+ 2 \, \ket{1,0} \otimes \ket{1,0}\Big) \,. \label{T20}
\end{align}\end{subequations}
In the case of spherical symmetry, the only non-zero component of $\Psi$ is along $\ket{2,0}$.
Recalling the normalisation difference with the basis given by $\tl\tau_i$, we can write
\equ\label{PsiSym}
\Psi= - \beta \big(\tl\tau_{+} \otimes \tl\tau_{-} + \tl\tau_{-} \otimes \tl\tau_{+}
+ 4 \, \tl\tau_{0} \otimes \tl\tau_{0}\big) \,.
\nequ
The minus sign is to have the same $\beta$ as in \cite{Krasnov1, Krasnov2}, where different conventions were used.

\medskip 
To write the equation \Ref{EqB}, we also need the identity in the spherical 
basis. The latter is the inverse $\tl\d^{ij}$ of \Ref{tildeTrace}, 
\equ\label{Id}
{\mathbbm 1} = \tl\d^{ij} \tl\tau_i\otimes\tl\tau_j 
             = \tl\tau_{0} \otimes \tl\tau_{0}
-\f12 \tl\tau_{+} \otimes \tl\tau_{-} -\f12 \tl\tau_{-} \otimes \tl\tau_{+} \,.
\nequ
This identity naturally belongs to the singlet ${\bf 0}$ and corresponds to the standard SU(2) scalar product $g^{mn}=(-1)^m \d_{m, -n}$, once the different
normalisation of the $\tl\tau_{\pm}$ is taken into account.

\medskip 
We now have all the ingredients to write the metricity equations \Ref{EqB}. 
Using the metric $\tl\d_{ij}$ in \Ref{tildeTrace} to take the scalar products, 
we compute 
\eqa
\Psi^i{}_k \Psi^{kj} \, \tl\tau_i \otimes \tl\tau_j &=& 
{\beta^2} (-2\tl\tau_{+} \otimes \tl\tau_{-} - 2\tl\tau_{-} \otimes \tl\tau_{+}
+ 16 \tl\tau_{0} \otimes \tl\tau_{0})\,, \\
\Psi^i{}_k \Psi^{k}{}_l \Psi^{lj} \, \tl\tau_i \otimes \tl\tau_j &=& 
-{\beta^3} (4\tl\tau_{+} \otimes \tl\tau_{-} + 4\tl\tau_{-} \otimes \tl\tau_{+}
+ 64 \tl\tau_{0} \otimes \tl\tau_{0}) \,,
\neqa
from which
$
z_1 = \Psi^i{}_j \Psi^{j}{}_i = 24 \beta^2$, 
$z_2 = \Psi^i{}_k \Psi^{k}{}_l \Psi^{l}{}_{j} = -48 \beta^3.$

\medskip 
Being both $z_1$ and $z_2$ functions of $\beta$, it follows that the 
modification to general relativity is characterised in the spherically 
symmetric sector by the functional $\Phi(\beta)$ of a single scalar function. 
As a consequence we also have that the partial derivatives are simply 
$\d_{z_1} \Phi = \pb / {48\beta}$, $\d_{z_2} \Phi = -\pb / {144\beta^2}$, 
where we have introduced the shorthand notation 
$ \pb = {\d \Phi}/{ \d \beta}$. Notice that $\Phi$ and $\Psi$ have the same 
dimensions, thus $\pb$ is dimensionless. 

\medskip 
Putting all this together, eq.~\Ref{EqB} gives 
\equ
B^i \w B^{j} \, \tl\tau_i \otimes \tl\tau_j = \f13 \, B^k \w B_k \, 
\left[ \left(\f14\pb-\f12\right) \big(\tl\tau_+ \otimes \tl\tau_- + \tl\tau_- \otimes \tl\tau_+ \big) +
(\pb+1)\, \tl\tau_0 \otimes \tl\tau_0 \right] \,.
\nequ
From this we can obtain the following five independent equations,
\begin{subequations}\begin{align}\label{equazioni}
& B^+ \w B^+ = B^- \w B^- = 0 \,, \\
& B^+ \w B^0 = B^- \w B^0 = 0 \,, \\
& 2 B^+ \w B^- + B^0 \w B^0  = \f12 \pb \, B^k \w B_k \,. 
% = \f12 \pb \,(B^0\w B^0 - 4 B^+\w B^-),
\end{align}
Using the scalar product \Ref{tildeTrace}, the last equation can be written as
\equ
\f{B^+ \w B^-}{B^0\w B^0} = \f{\pb-2}{4(1+\pb)} \,.
\nequ
\end{subequations}

\medskip 
Proceeding as discussed above in Section~\ref{SecSolving}, we now look for 
a solution in the form 
\equ\label{ansatz}
B^i = D^i{}_j(c_n) \wtl\Sigma^j(e) \,,
\nequ
with $\wtl\Sigma^i(e)$ given by \Ref{tildeSigma}, and 
$c_n$ five parameters to be related to $\Psi^{ij}$ through \Ref{EqB}.
In the spherical
basis, a generic linear combination that preserves self-duality is given by 
(see e.g. Appendix of \cite{Krasnov1})
\equ
D(c_n) = \left( \begin{array}{ccc} c_0 & 0 & 0 \\ 0 & c_+ & c_{+-} \\ 0 & c_{-+} & c_- \end{array}\right) \,.
\nequ
Plugging this into \Ref{equazioni}, we have 
\equ
c_+ \, c_{+-} = 0, \qquad  c_- \, c_{-+} = 0, \qquad
\f{ c_+ \, c_{-} + c_{+-} \, c_{-+} }{-2 (c_0)^2} =  \f{\pb-2}{4(1+\pb)}.
\nequ
A simple parametrisation of the solution is given by $c_{+-}=c_{-+}=0$, $c_0=1$ and $c+=c-\equiv c$ with
\equ\label{c}
c^2 = \f{2-\pb}{2(1+\pb)}.
\nequ

\medskip
In conclusions, the $B^i$ field given by the triple of 2-forms
\equ\label{Bsph}
B^+= {c}\, n\w \mb, \qquad B^-= {c} \, m\w l, \qquad B^0=l\w n - m\w\mb,
\nequ
solves the metricity equations \Ref{EqB} and the reality conditions,
and it is self-dual with respect to the metric \Ref{gNP}. 
Note that although $c^2$ might be negative, hence $B$ complex,
\Ref{gNP} is always real and Lorentzian (up to singularities).
Note also that the relation between $B^i$ and the metric \Ref{gNP} breaks 
down when $\pb=2$ or $\pb=-1$. These situations can indeed arise, and are 
the non-metric singularities which we will discuss below. 

\medskip 
Needless to say, having a metric at disposal is instrumental for the physical interpretation of the theory.
The strategy is to use \Ref{Bsph} to write the remaining field equations as differential equations for
$f$, $g$ and $\beta$. Once the theory has been reformulated in such more familiar terms, also the effect of the
modification become more transparent.

\medskip
Thus far, the only place where we used spherical symmetry was the fact that 
$\Psi$ has the single component \Ref{PsiSym}. 
(In the general case the equations \Ref{equazioni} have non-vanishing right 
hand sides and depend on the whole five components of $\Psi$, which get 
related to the coefficients $c_n$). 
We now specialize the Newman-Penrose tetrad, and thus the solution \Ref{Bsph} 
for the $B$ field, to the case of spherical symmetry. We choose spacetime 
coordinates as follows:  
\equ\label{NPsph}
l = \f1{\sqrt{2}} \left(f \,dt- g \, dr\right), \qquad
n = \f1{\sqrt{2}} \left(f \,dt+ g \, dr\right), \qquad
m = \f{R}{\sqrt{2}} \left(d\th + i \, \sin \th \, d\phi \right), 
\nequ
and $\bar m = m^*$. Through eq.~\Ref{gNP}, they define the metric
\begin{eqnarray}
\label{gsph}
ds^2  = f^2dt^2 - g^2 dr^2 - R^2(d\th^2 + \sin^2 \th \, d\phi^2 ) , \qquad e= fg R^2\sin\th,
\end{eqnarray}
which can be recognised as the standard ansatz for a spherically symmetric 
spacetime. We initially take $f(t,r)$, $g(t,r)$ and $R(t,r)$ functions of time as well as the radius.

\medskip 
Using the tetrad \Ref{NPsph}, we can write the components $B^i$ in \Ref{Bsph} 
in the following matricial form, 
\equ\label{Bmatrix}
B^0_{\mu\nu} = \left(\begin{array}{cccc}
0 & f g & 0 & 0 \\
 & 0 & 0 & 0 \\
& & 0 & i R^2 \sin \th \\
&& &0 
\end{array}\right), \qquad
B^{\pm}_{\mu\nu} = \f{c}2 \left(\begin{array}{cccc}
0 & 0 & \pm R f & -i R  f \sin \th  \\ 
 & 0 & R g & \mp i R g \sin \th  \\
& & 0 & 0 \\
&&&0 
\end{array}\right) \,.
\nequ

%%%%%%%%%%%%% 
\subsection{Maxwell fields} 
%%%%%%%%%%%%%
With the explicit form of $B^i$ at hand, we can immediately compute 
the Maxwell stress-energy tensor \Ref{MT}. 
It is straightforward to show that when a Maxwell field respects 
the spherical symmetry, the field strength takes the form 
\equ
F(A)= E(t,r) \, l \w n + B(t,r) \, m \w \mb 
\nequ 
with two scalars, $E(t,r)$ and $B(t,r)$, corresponding respectively 
to the electric and magnetic field. Then, it follows from the source-free 
Maxwell equations $dF=0$ and $d*F=0$ (which are still valid in the modified 
theory in terms of the derived metric) that 
$B(t,r)= const.  \times R(t,r)^{-2}$, $E(t,r)= const. \times R(t,r)^{-2}$. 
For simplicity, hereafter we restrict our attention to the case $B(t,r)=0$, 
and thus in units $4\pi\eps_0=1$,  
\equ\label{FM} 
F(A)= \frac{Q}{R(t,r)^2} \, l \w n \,, 
\nequ
where the constant $Q$ is chosen to be the conserved electric charge 
defined by 
\equ \label{def:charge}
 Q = \frac{1}{4\pi}\int_{S^2} \star F \,,    
\nequ
on a $2$-sphere $S^2$.

\medskip 
To evaluate eq.~\Ref{MT} we need the self and antiself-dual parts of \Ref{FM}, 
which can be computed from \Ref{FBB} and the explicit form \Ref{Bmatrix} 
of $B$. The only non-vanishing components turn out to be 
\equ\label{FMsph}
F^{(\pm)}_{tr} = \f{Q fg}{2 R^2}, \qquad F^{(\pm)}_{\th\phi} = \pm \, 
{i } \, \f{Q}{2} \sin\th \,.
\nequ
From the equation of motion \Ref{eqphi} we have
\equ
F^{(+)}_{\mu\nu}=\vp_i\, B_{\mu\nu}^i = \vp^0 \, B^0_{\mu\nu}
-2\, \vp^+ \, B^-_{\mu\nu}-2\, \vp^- \, B^+_{\mu\nu} \,,
\nequ
which gives
\equ\label{phiSph}
\vp^0 = \f{Q}{2 f g R^2}, \qquad \vp^+ = \vp^- =0 \,,
\nequ
Finally, plugging \Ref{FMsph} and \Ref{phiSph} into \Ref{MT}, 
the only non-zero components of the energy-momentum 2-form compute to
\equ\label{MTsph}
 \tau_{tr}^0 =  - \f{Q^2 fg }{8\pi R^4}, \qquad 
 \tau_{\th\phi}^0 =i \f{Q^2}{{8\pi}R^2}\sin \theta \,. 
\nequ

%----------------------------------------------------------------------
\subsection{Cartan equations}\label{SecCart2}
%----------------------------------------------------------------------
The next step is to solve the Cartan equation \Ref{EqCart2}, using the explicit form of $B^i$ in \Ref{Bmatrix},
to compute the connection $\om$ in terms of the metric and $\beta$.
Since we are using the spherical basis, we have to adapt the covariant derivative. Recalling its general definition
$d_\om B=d\om + [\om,B]$ and using the commutators in the $\tl\tau_i$ basis, we get
\begin{align}\label{Cart2Sph}
& dB^0 +2 \Big( \om^- \w B^+ - \om^+ \w B^- \Big) = 0 \,,
& dB^{\pm} \pm \Big(\om^0 \w B^\pm - \om^\pm \w B^0 \Big) = 0 \,.
\end{align}

\medskip
To solve the equations we proceed as follows. First, we use \Ref{Bmatrix} and 
write down the non-zero components of the 3-forms $dB^i$. 
Then, we take the wedge product of the equations \Ref{Cart2Sph} with all four 
basis vectors $l, \ldots, \mb$, and read off the components of $\om$ in this 
basis, as functions of the metric and $\beta$. 
The procedure is straightforward but space consuming, so we
report here only the final result (more details can be found in \cite{Krasnov1,Krasnov2}),
\eqa\label{omega}
\om^\pm = - \f12 P_\pm (d\th \mp i\, \sin\th \, d\phi) \,, \qquad
\om^0 = F \, dt + G \,dr - i\cos\th\, d\phi \,,
\neqa
where we have introduced the following shorthand notation,
\equ
P_\pm = \f{R'}{g_*}  \mp \f{\dot R}{f_*} \,, \qquad
F = \f{(R f_*)'}{R g_*} - \f{R' f_*}{ c^2 R g_*} \,, \qquad
G = \f{(R g_*)^{\cdot}}{R f_*} - \f{\dot{R} g_*}{c^2 R f_*} \,, 
\nequ
with $f_*=c f$ and $g_*=c g$. 
Here and in the following $\dot{}$
and ${}'$ denote, respectively, the derivative by $t$ and $r$.  

\medskip 
The curvature in the spherical basis is
\begin{align}\label{FSph}
& F^0 = d\om^0 + 2 \om^- \w \om^+ \,,
& F^\pm = d\om^{\pm} \pm \om^0 \w \om^\pm \,,
\end{align}
where from \Ref{omega} we have
\begin{subequations}\label{Fsph}
\begin{align}
&\p_\mu \om_\nu^0 = \left(\begin{array}{cccc}
\dot{F} & \dot{G} & 0 & 0 \\
F' & G' & 0 & 0 \\
0 & 0 & 0 & i\,s_\th \\
0&0&0&0 
\end{array}\right), &
\p_\mu \om_\nu^\pm = \f12 \left(\begin{array}{cccc}
0 & 0 & -\dot{P}_\pm & \pm i\,s_\th\, \dot{P}_\pm \\
0 & 0 & -P_\pm' & \pm i\,s_\th\, P_\pm' \\
0 & 0 & 0 & \pm i\,c_\th\, P_\pm \\
0&0&0&0 
\end{array}\right), 
\\
&\om^-_\mu \, \om_\nu^+ = \f{P_+\, P_-}{4} \left(\begin{array}{cccc}
0 & 0 & 0 & 0 \\
0 & 0 & 0 & 0 \\
0 & 0 & 1 & -i\,s_\th \\
0&0& i \, s_\th & s_\th^2 
\end{array}\right) \,, &
\om^0_\mu \, \om_\nu^\pm = \f{P_\pm}{2} \left(\begin{array}{cccc}
0 & 0 & -F & \pm i \, s_\th \, F \\
0 & 0 & -G & \pm i \, s_\th \, G \\
0 & 0 & 0 & 0 \\
0 & 0 & i\,c_\th & \pm c_\th \, s_\th
\end{array}\right) \,.
\end{align}
\end{subequations}
Here $s_\th=\sin\th$ and $c_\th=\cos\th$.

%----------------------------------------------------------------------
\subsection{Field equations}\label{SecFE}
%----------------------------------------------------------------------
Using \Ref{PsiSym} and \Ref{FSph}, the field equations \Ref{EqF2} in the spherical basis read 
\begin{subequations}\label{EqFApp}
\eqa\label{EqFAppA}
d \om^\pm \pm \om^0 \w \om^\pm + (\beta + \Phi) \, B^\pm &=& 8\pi\GN \, \tau^\pm \,,  \\ 
\label{EqFAppB}
d \om^0 + 2 \om^- \w \om^+ + (\Phi-2\beta) \, B^0 &=& 8\pi\GN \, \tau^0 \,. 
\neqa
\end{subequations}
Here $\om=\om(B,\beta)$ through \Ref{omega} and $\Phi=\Phi(\beta)$ as seen above.
For a spherically symmetric configuration of Maxwell's field,  
we found that the only non-zero components of the energy-momentum tensor 
are $\tau^0_{tr}$ and $\tau^0_{\th\phi}$. 
Using also the explicit expressions 
\Ref{Bmatrix} and \Ref{Fsph}, the non-vanishing components of equations \Ref{EqFAppA} are
\begin{subequations}\label{Eq:}\begin{align}
&  \dot{P}_\pm \pm {P_\pm}\, F \mp (\beta+\Phi) R f_* = 0\,,  
\label{Eq:a} 
\\ 
\label{Eq:b} 
&  P_\pm' \pm P_\pm \, G - (\beta+\Phi) R g_* = 0  \,,
\end{align}
and those of \Ref{EqFAppB}
\begin{align}\label{Eq:c}
& \dot{G} - F' + (\Phi-2\beta) \f{f_* g_*}{c^2} = 8\pi\GN \, \tau^0_{tr} \,, 
\\ 
\label{Eq:d}
& 1- P_+ P_- + R^2 (\Phi-2\beta) = 8\pi\GN \, \f1{i \sin \th } \, \tau^0_{\th\phi} \,.
\end{align}\end{subequations}
Among the eight equations \Ref{Eq:}, only three are independent because of the Bianchi identities \cite{Krasnov2}.
Below in Section \ref{SecStatic} we write the three independent equations in the form suggested in \cite{Krasnov2}. 
Before doing so, we show the validity of part of the Birkhoff theorem 
that any spherically symmetric vacuum solutions are static.   

%%%%%%%%%%%%%
\subsection{Birkhoff's Theorem}
\label{sec:Birkhoff}
%%%%%%%%%%%%%

In \cite{Krasnov2}, Krasnov and Shtanov proved the Birkhoff's theorem that the 
spherically symmetric solutions of this theory are necessarily static for 
the vacuum case. 
Their proof is slightly unconventional, so we find it useful to reproduce the 
same results for our case, following a more standard procedure~\cite{Hawking}. 
We assume that $R(t,r)$ be not constant, $d R \neq 0$. 
This, in particular, implies that we can choose the gauge so that $R(r,t)=r$. 
Then, the field equations above reduce to 
\equ\label{R=r}
P_\pm = \f1{g_*}, \qquad F= \f{{f_*'}}{g_*}+\f{{f_*}}{r g_*}\left( 1-\f1{c^2}\right), \qquad G=\f{\dot{g_*}}{f_*} \,.
\nequ 
The first two of equations \Ref{Eq:} simplify to
\begin{subequations} \label{B}
\begin{align}
& \label{B1} -\f1{g_*^2} \, \dot{g_*} \pm \f1{g_*} F \mp (\beta+\Phi) r f_* 
= 0 \,, \\
\label{B2}
& -\f1{g_*^2} \, g_*' \pm \f{\dot{g_*}}{f_* g_*}\, G - (\beta+\Phi) r g_* = 0
\,.
\end{align}\end{subequations} 
We now show that the solution is necessarily static, i.e. that $\dot{f}=\dot{g}=\dot{c}=0$.
First of all, summing the two equations in \Ref{B1} gives $\dot{g_*}=0$, 
thus $G=0$ from the third of \Ref{R=r}. 
It then follows from \Ref{B2} that ${\beta}+{\Phi}$ is a function of 
only $r$ and thus $(1+\Phi_\beta){\dot \beta}=0$. 
As we briefly mentioned above $\Phi_\beta =-1$ corresponds to 
the singular case, which we are not going to consider here. 
We therefore conclude that $\beta$ itself is $t$-independent and 
so are $\Phi$ and $c^2$. 
Subtracting the two equations \Ref{B1} gives $ F/{f_*} = r g_* (\beta + \Phi)$, implying that $(F/f_*)^\centerdot=0$. 
Then, it follows from \Ref{R=r} and $(F/f_*)^\centerdot=0$ that 
$(f_*'/f_*)^\centerdot=0$, namely $f_*'/f_*=a(r)$ with $a(r)$ being some 
function of $r$; 
the latter equation allows for a time-dependent integration constant, 
$f_*(t,r)= b(t) \int a(r) dr$, but this is of the type that can always be 
absorbed by a change of coordinates $t\mapsto \tau$ with $d\tau = b(t) dt$. 

\medskip 
Note that besides this staticity property, Birkhoff's theorem also asserts 
that such a spherically symmetric static metric, if satisfying the vacuum 
Einstein equations, is necessarily locally isometric to the Schwarzschild 
metric~\cite{Hawking}. For our case, in order to have the corresponding 
assertion, we have to integrate eqs.~(\ref{Eq:c}) and (\ref{Eq:d}) 
with the stress-tensor for non-vanishing Maxwell fields. We perform this 
integration only numerically and do not have analytic solutions. 
For this reason we do not have the corresponding assertion in the same 
decisive level as the original Birkhoff's theorem for the vacuum 
Einstein gravity.

%---------------------------------------------------------------------
\subsection{Equations of motion for static fields}\label{SecStatic}
%---------------------------------------------------------------------
Using the staticity and the choice of gauge $R(r)=r$, 
the equations reduce to
\begin{subequations} \label{EqSph}
\begin{align}\label{EqSph1}
& \beta+\Phi = \f{1}{r f_* g_*} F = -\f{g_*'}{r g_*^3} 
\\\label{EqSph2}
& -F' + (\Phi-2\beta)\f{f_* g_*}{c^2} = 8\pi\GN \, \tau^0_{tr}   
\\\label{EqSph3}
& 1- \f1{g_*^2} + r^2(\Phi-2\beta) = \f1{i\sin\th} \, 8\pi\GN \, \tau^0_{\th\phi} 
\end{align}
\end{subequations} 
Here \Ref{EqSph3} is an algebraic relation, and \Ref{EqSph1},
\Ref{EqSph2} are differential equations for $f_*$, $g_*$ and $\beta$, of which only two are independent thanks to the Bianchi identities.
Following \cite{Krasnov2}, we now reexpress these equations in a more convenient form.
First of all, \Ref{EqSph1} and the second of \Ref{R=r} immediately imply that
\equ\label{pippo}
\f{(f_* g_*)'}{f_* g_*} = -\f1r\left(1-\f1{c^2} \right) \,.
\nequ
Secondly, taking the derivative of  \Ref{EqSph2} and using the above equation, we obtain
\eqa\label{pluto}
\beta'(2-\pb) &=& -\f2{r}\left(3\beta + \f{c^2}{f_* g_*} \, 
8\pi\GN \,\tau^0_{tr}\right) \,.
\neqa
The three equations \Ref{EqSph3}, \Ref{pippo} and \Ref{pluto} are the field equations of the theory
in the spherically symmetric sector. They give dynamics for $f_*$, $g_*$ 
and $\beta$, and thus for the metric
\Ref{gsph}. 
For $Q=0$ these equations reduce to Eq. (35) of \cite{Krasnov2}.
Finally, using the explicit expressions of the energy-momentum tensor \Ref{MTsph} and of $c^2$ (eq.~\Ref{c}), and defining $x(r) \equiv f_*(r) g_*(r)$, 
we have the independent equations 

\begin{subequations} \label{MaxEq}
\eqa \label{MaxEq1}
\f{x '}{x} &=& \f{3 \pb}{r\left(2-\pb \right)} \,, 
\\ \label{MaxEq2}
(2-\pb)\beta' &=& -\f{6\beta}{r} 
                  + \f{2G_N Q^2}{r^5} \,, 
\\ \label{MaxEq3}
\f1{g_*^2} &=& 1 - (2\beta-\Phi) r^2 - \f{\GN Q^2}{r^2} \,.
\neqa\end{subequations}

Before proceeding, let us check that for constant $\Phi=- \Lambda/3$ 
we recover the standard results of the general relativity case.

%%%%%%%%
\subsection{Recovering the standard Reissner-Nordstr\o m metric} 
%%%%%%%%
When $\pb=0$ we have $c^2=1$ and, from \Ref{MaxEq1}, that $f_* g_* = fg $ is a constant.
Taking the latter to be 1, $f=1/g$.
We are left with a single differential equation, from \Ref{MaxEq2},
\equ
\beta'= -\f{3\beta}{r} + \f{\GN Q^2}{r^5} \,.
\nequ
This equation is solved by
\equ\label{betaGR}
\beta(r) = \f{\GN M}{r^3} - \f{\GN Q^2}{r^4} \,,
\nequ
where $M$ is the integration constant. 
Finally from \Ref{MaxEq3} we read
\equ\label{grr}
\f1{g^{2}} = 1 - \f{2 \GN M}{r} + \f{\GN Q^2}{r^2} -\f13 \Lambda r^2
\nequ
which can be immediately recognised to give the $g_{00}= g^{-2}$ component of 
the Reissner-Nordstr\o m metric with cosmological constant $\Lambda$.

%%%%%%%%
\subsection{Metric singularities}
%%%%%%%%
The field equations \Ref{MaxEq} give the dynamics for $f_*$, $g_*$ and $\beta$, in terms of which
the metric \Ref{gsph} reads
\equ\label{g00}
ds^2 = \f{2(1+\pb)}{2-\pb} \left[ {f_*^2} \, dt^2 - g_*^2 \, dr^2 \right] - r^2 \, d^2\Omega \,.
\nequ
Notice the possibility of singularities if values of $r$ exist such that $\pb=2, -1$. These are the 
\emph{non-metric} singularities found and discussed in \cite{Krasnov2}. 
As shown above in Section~\ref{sect:constraints}, these are points where the 
relation between $B$ and the metric breaks down. To understand the theory, 
it is useful to distinguish the following three types of apparent singularities 
which can occur: 
\begin{itemize}
\item Points where $g_*^2(r)$ diverges and $x(r)$ is finite are Killing horizons, the outermost of which may be viewed as the black hole (event) horizon, depending upon the asymptotic structure. 
\item Points where $\pb=2$. We call this non-metric singularities of type (1).
\item Points where $\pb=-1$. We call this non-metric singularities of type (2).
\end{itemize}
The nature of these non-metric singularities will be discussed below, with explicit solutions at hand.

\medskip
It is at this point important to comment on the conformal ambiguity present in the theory.
As discussed earlier in Section \ref{SecSolving}, the procedure to extract a metric is defined up to
a conformal transformation, and in the electrovacuum theory there is no principle which would
select a physical metric in the conformal class. In particular,
multiplying \Ref{g00} by a conformal factor generically depending on $\beta$
gives an equally valid line element in this modified theory of gravity. 
A possible mechanism to distinguish a preferred metric has been studied in \cite{KrasnovSmall}
by introducing test particles and demanding that they follow geodesics. 

\medskip
Since we restrict our analysis to the coupling to electromagnetism and Maxwell's equations are conformally invariant, the ambiguity is present in our case, and we have to arbitrarily select a metric in order to
investigate its singularities. Thus in
the absence of a physical criterium to distinguish a preferred metric, 
we choose to work with the the line element \Ref{g00},
which emerged naturally in our construction.\footnote{Another natural choice considered in the literature
\cite{Krasnov2} is to work with the metric such that the volume form coincides with $B^k \w B_k / 3$. 
This amounts to multiplying \Ref{g00} by $(1+\pb)^{-1/2}$.}

%%%%%%%%
\section{Choices of $\Phi$}\label{SecPhi}
%%%%%%%%
Thus far we have kept $\Phi$ arbitrary, without worrying about the origin of 
its non-constancy. Different choices of $\Phi$ have an impact on the classical 
behaviour of gravity in different ways and at different scales, allowing one 
to entertain the possibility of explaining in this way current puzzles, such 
as anomalous rotational curve in spiral galaxies, without appealing to dark 
matter scenarios. To that end, tailored profiles of $\Phi$ have been 
investigated in \cite{Krasnov2,Krasnov5}.

\medskip 
In introducing this modification of gravity, Krasnov pointed out that it could be seen as an effective
action coming from a model of quantum gravity where the fundamental field is not the metric, but rather
the 2-form $B$. Recall that $\Phi$ has length dimensions $[\Phi]=[\Psi]=-2$, and that
it is a function of $z_1={\rm Tr}\, \Psi^2$ and $z_2={\rm Tr}\, \Psi^3$ 
with dimensions $[z_1]=-4$ and $[z_2]=-6$, we see that a non-trivial
dependence of $\Phi$ on $\Psi$ introduces coupling constants with non-zero 
length dimensions. 
A natural coupling constant with such dimensions is the Planck length,
which leads us to postulate a quantum mechanical origin of a non-constant 
$\Phi$. At present, it is not clear what quantum gravity model could lead to 
such an effective action, but the use of $B$ instead of the metric as the 
fundamental field suggests to look at spin foam models as possible candidates.

\medskip 
In this paper we do not attempt to quantize the theory, we are only 
interested in classical consequences of different choices of $\Phi$. 
Consider for instance the case in which $\Phi$ is `analytic' near the origin 
in the sense that it can be Taylor expanded as 
\equ\label{expP}
\Phi(\Psi) = \Phi_0 + a_1 \, {\rm Tr}\, \Psi^2 + a_2 \, 
                             {\rm Tr}\, \Psi^3 + \ldots  
\nequ
As $[\Psi]=-2$, it follows that $[a_n]={2n}$, e.g 
$a_n = b_n \, (\hbar \GN)^n$ with $b_n$ dimensionless. 
Lack of control on the possible origin and on the quantum properties of 
the theory makes it hard to restrict the admissible forms of $\Phi$, and 
a priori we have no reasons to demand analyticity in the sense of \Ref{expP}. 
On the other hand, by studying the implications of a non-constant $\Phi$ at 
the classical level we can hopefully obtain restrictions on the admissible 
forms of $\Phi$.

%%%%%%%%
\subsection{On asymptotic flatness}\label{sect:AF}
%%%%%%%%
A useful requirement is that in the modified theory, there exist 
such solutions that approach at large distances an asymptotically flat 
solution of general relativity. 
This can put restrictions on the admissible choices of $\Phi$. 
Taking $\Lambda=0$ for simplicity, we demand that 
\eqa \label{condi:flat:c}  
c^2 &\rightarrow& 1 \,, 
\\
\label{condi:flat:fg}  
f^2, \, \, g^2 &\rightarrow& 1 + O(1/r) \,,  
\\
\label{condi:flat:beta}  
\beta &\rightarrow& 0 \,, 
\neqa
in the limit of large area radius $r \rightarrow \infty$. 
Since $c^2= ({2- \Phi_\beta})/{2(1 + \Phi_\beta)}$, the first of the 
asymptotic flatness condition, eq.~\Ref{condi:flat:c}, 
implies that $\Phi_\beta$ cannot approach a non-vanishing constant, and 
hence that with eq.~\Ref{condi:flat:beta}, $\Phi$ cannot be linear to $\beta$. 
Therefore the asymptotic flatness conditions above require that 
if Taylor-expanded in $\beta$, 
$\Phi$ must start from at least the quadratic order of $\beta$:  
$\Phi \propto \hbar \GN \beta^2 + O(\beta^3) $. 
This also is a justification of our discussion about the form, 
eq.~\Ref{expP}, from a physical view point. 
The simplest modification to general relativity is thus to truncate 
the expansion \Ref{expP} to the first non-trivial term, 
$\Phi=\Phi_0 + a_1 \, {\rm Tr}\, \Psi^2 = \Phi_0 + 24 a_1 \beta^2$. 
(When $\Phi_0 \neq0$, the solution would be asymptotically (anti-)de Sitter.) 
This quadratic case will be discussed in detail in the subsequent sections.

\medskip 
Since our present metric is static, admitting the time-like Killing vector 
field $\xi^\mu = (\partial/\partial t)^\mu$, the ADM mass may be given by 
the Komar integral on $2$-sphere, ${S^2_\infty}$, at spatial infinity, 
\equ \label{def:ADMmass}
 M = - \f{1 }{8\pi \GN } \int_{S^2_\infty} \star (\nabla \xi ) \,.
\nequ
(Note that the conserved electric charge, $Q$, is always given 
by \Ref{def:charge}, irrespective to the conditions, eq.~\Ref{condi:flat:fg}.)
Since 
\equ
 f^2 = \frac{x^2}{c^2g_*^2} 
     = x^2 \left[1 -(2\beta - \frac{A}{2}\beta^2)r^2 + O(1/r^2) \right] \,, 
\nequ 
the second of eq.~\Ref{condi:flat:fg} implies that 
\equ
  x^2 \rightarrow  1 + O(1/r) \,, \quad 
  \beta \rightarrow \frac{C}{r^3} + O(1/r^4) \,,  
\nequ
with some constant $C$. 
(In fact, integrating eq.~\Ref{MaxEq1} at asymptotic region implies 
$x^2 \sim \exp[-(AC)/r^3] \sim 1+ O(1/r^3)$.) 
Then, the ADM mass---which is now written 
$ M  = {r^2c^2}(f^2)'/{2x}$---is well-defined and becomes, $C=\GN M$, 
in accordance with the standard GR results, eq.~\Ref{betaGR}.

%%%%%%%%
\subsection{Linear modification}\label{sec:linear}
%%%%%%%%
When $\Phi$ is linear in $\beta$, then $x$ would either diverge or vanish 
at $r \rightarrow \infty$. Therefore solutions in a modified theory linear in 
$\beta$ would not be asymptotically flat. Since linearity in $\beta$ can only 
happen if $\Phi(\Psi)$ is non-analytic at the origin in the sense of the 
expansion eq.~\Ref{expP}, one may ask whether the lack of asymptotically flat 
static solutions characterises all ``non-analytic'' choices of $\Phi$. 
This would be a rather useful restriction on the admissible choices of $\Phi$.
We leave this question open for future investigations.

\medskip 
Nevertheless, the linear modified case can be solved easily, and 
thus it is instructive to consider this case first.  
Let us take $\Phi=\Phi_0+a \beta$, with $a$ being some 
dimensionless parameter. In this case $\pb$ is constant and eq.~(\ref{c}) 
reads  
\begin{equation} 
c^2=\f{{2-a}}{2(1+a)}.
\end{equation}
The constancy of $c^2$ trivialises some of the non-linearities of the field 
equations, thus significantly simplifying the study of the spacetime 
structure. 
Assuming $a\neq 2, -1$, the field equations are 
\begin{subequations} \label{MaxLin} \eqa \label{MaxLin1}
\f{x'}{x} &=& \f{3 a}{\left(2-a \right)r} \,, 
\\ 
\label{MaxLin2}
(2-a) \beta' &=& -\f{6\beta}{r} + \f{2 \GN Q^2}{r^5}\,, 
\\ \label{MaxLin3}
\f1{g_*^2} &=& 1 + \Phi_0 r^2 - (2-a)\beta r^2
                 - \f{\GN Q^2}{r^2} \,.
\neqa\end{subequations}
The equation \Ref{MaxLin1} can be easily solved to give
\equ
x(r) = c_1 \, r^{\f{3a}{2-a}} = c_1 \, r^{(1-c^2)/c^2} \,,   
\nequ
with $c_1$ being an integration constant. 
From \Ref{MaxLin2} we obtain
\equ\label{linexp}
\beta(r) = c_2 \, r^{-\f{6}{2-a}} 
         + \f{\GN Q^2}{(2a-1)} \, r^{-4} \,, 
\nequ
where $c_2$ is another integration constant. The integration constants can be fixed to reproduce the standard GR result 
\Ref{betaGR} for $a=0$, specifically $c_1=1$, $c_2=\GN M$. Then we have 
\equ
 \f1{g_*^2} = 1 + \Phi_0 r^2 - \f{(2-a)\GN M}{r^{1/c^{2}}} 
            + \f{3(1+a)}{2a-1}\f{\GN Q^2}{r^2} \,.
\nequ 
Note the different asymptotic behaviour from the standard 
Reissner-Nordstr\o m solution, and in particular the absence of asymptotic 
flatness. 
This means, in particular, that $M$ cannot be interpreted as the ADM mass. 
Note how the location of the zeroes of $1/g_*^2$ depends also on $a$. 
In the next Sections, when we treat the quadratic modification, we will see 
that a similar modification of the horizon structure happens within solutions 
which are asymptotically the Reissner-Nordstr\o m metric.

%%%%%%%%
\section{Modified self-dual gravity with Maxwell field: Quadratic case}
\label{sec: MG: Quadratic}
%%%%%%%%
In the rest of the paper we focus on the quadratic modification 
$\Phi(\beta) = \Phi_0 + 24 a_1 \beta^2$. 
For commodity, we redefine $a_1= \hbar \GN A / 48$, and from now on we work 
in units $\hbar=\GN=1$. 
We then have $\Phi(\beta) = \Phi_0 + A \beta^2 / 2$, $\pb = A \beta$ and
\equ
c^2 = \f{{2-A\beta}}{2(1+A\beta)} \,,
\nequ
which is positive for $A\beta\in (-1,2)$ and negative otherwise.

\medskip 
As a consequence of this choice of $\Phi$, we expect larger departures 
from general relativity in regions of larger curvature. 

%%%%%%%%
\subsection{Nariai-Bertotti-Robinson solutions}
%%%%%%%%
Our general formulation of the field equations \Ref{Eq:} allows us to look for 
solutions with $R(t,r)=R_0$ constant. Such solutions exist in general 
relativity: They are not asymptotically flat and have a Cartesian product 
structure. The simplest case is the neutral Nariai metric \cite{Nariai}, 
a solution to the Einstein equation with a positive cosmological constant 
given by a direct product $dS_2\times S^2$. Setting $R(r,t)=R_0$ in \Ref{Eq:}, 
we have 
\equ\label{R0}
P_\pm = 0, \qquad F= \f{{f_*'}}{g_*}, \qquad G=\f{\dot{g_*}}{f_*} \,.
\nequ 
Equations \Ref{Eq:a} and \Ref{Eq:b} reduce to
\equ\label{Nariai1}
(\beta+\Phi(\beta))f_* = (\beta+\Phi(\beta))g_* = 0 \,,
\nequ
whereas \Ref{Eq:d} reduces to
\equ\label{Nariai2}
1 + R_0^2 \, (\Phi(\beta)-2\beta) = 0 \,.
\nequ
Avoiding the unphysical case of $f_*$ and $g_*$ everywhere vanishing, 
equations \Ref{Nariai1} give an algebraic constraint for $\beta$.  
Thus, the solutions to the constraint $\beta + \Phi(\beta)=0$ will 
have constant Weyl curvature. 

\medskip 
In general relativity, i.e., $\Phi= \Phi_0=-\Lambda/3$, the constraint is 
$\beta=\Lambda/3$, which plugged into eq.~\Ref{Nariai2} gives 
\equ\label{PNariai} 
R_0^2 = \f1{\Lambda} \,.
\nequ
This fixes the value of $R_0$ dynamically, 
and the Nariai solution follows from \Ref{Eq:c}. 

\medskip 
In the quadratically modified case, the constraint is 
\equ\label{Bert}
\f12 A \beta^2 + \beta -\f13 \Lambda= 0 \,,
\nequ
from which $\beta= (-1\pm \sqrt{1+(2/3)\Lambda A})/A$. 
Substituting this value into \Ref{Nariai2} we have 
\equ\label{PNariaiMod}
R_0^2 = \frac{A/3}{\pm \sqrt{1+(2/3)\Lambda A}-1} \,. 
\nequ
Therefore the Nariai solution still exists, but with a different value of its 
constant radius. This new value now depends on the modification parameter $A$, 
as well as the cosmological constant $\Lambda$. For $A>0$, we take the 
solution of $\beta$ with positive sign, and $R_0^2$ in \Ref{PNariaiMod} is 
positive provided $\Lambda$ is positive, as in \Ref{PNariai} above. 
For $A<0$, the solution with positive sign again requires $\Lambda>0$, 
but the solution with negative sign allows any $\Lambda<3/2|A|$. 
This, in particular, implies that the Nariai metric of $dS_2 \times S^2$ 
can be a solution even when $\Lambda <0$ if $A<0$ with negative sign of 
eq.~\Ref{PNariaiMod}. 
It is however only the solutions with positive sign and $\Lambda>0$  
that reduces to eq.~\Ref{PNariai} in the limit $A\mapsto 0$, 
i.e., the limit to the standard general relativity case. 

\medskip 
Another case is the Bertotti-Robinson metric \cite{Bertotti,Robinson}, a 
solution to the Einstein-Maxwell equations with vanishing cosmological 
constant, given by the direct product of $AdS_2 \times S^2$ with 
each 2-dimensional manifold having the same constant curvature radius. 
This metric has vanishing Weyl curvature and therefore is still a solution 
to the modified theory. Indeed, taking $\Lambda=0$ in \Ref{Bert} we see that $\beta=0$ is a solution
also for $A\neq 0$.

%%%%%%%%
\subsection{Perturbative analysis of charged black holes} 
%%%%%%%%

Let us now look for asymptotically flat solutions. Fixing $R(r)=r$,
the equations \Ref{MaxEq} take the form 
\begin{subequations}\label{MaxQ}\eqa  \label{MaxQ1}
\f{x'}{x} &=& \f1r \, \f{3A \beta}{2-A\beta} \,, 
\\  
\label{MaxQ2}
\beta'(2-A\beta) &=& -\f{6\beta}{r}+\f{2 Q^2}{r^5} \,, 
\\ 
\label{MaxQ3} 
\f1{g_*^2} &=& 1 - r^2(2\beta-\f12A\beta^2-\Phi_0) - \f{Q^2}{r^2} \,. 
\neqa\end{subequations}
With respect to general relativity, two immediate effects of the
modification are visible:
\begin{itemize}
\item From \Ref{MaxQ1}, $x=$const is not anymore a solution, so we do not 
have the familiar form, $f^2=1/g^2$, of the spherically symmetric 
Einstein-Maxwell system. 
\item In eq.~\Ref{MaxQ2} the term proportional to $A$ makes the differential 
equation non-linear. 
\end{itemize}
The non-linearity of the system is still manageable for $Q=0$, the case that 
was solved exactly in \cite{Krasnov2}, and which we review below. However for 
$Q\neq 0$ the situation is more involved, and unfortunately the system eluded 
our attempts to find exact analytic solutions. 
A solution can be found perturbatively in the regime $|A\beta|\ll 1$, giving
\eqa\label{pert}
x(r) &=& 1 + A \left( -\f{M}{2r^3} + \f{3 Q^2}{8 r^4}\right) + o(A^2) \,,
\\
\label{betapert}  
\beta(r) &=& \f{M}{r^3} - \f{Q^2}{r^4} 
          + \f{A}{4r^6} \left(M - \f{Q^2}{r}\right)^2 + o(A^2) \,, 
\neqa 
where we fixed the integration constants to match general relativity at zeroth 
order in $A$. A posteriori we see that the approximation is valid for 
$r\gg (AM)^{1/3}$. 
The perturbative solution is useful to fix the integration constants, 
because it turns out that, unlike the linear modification (cf. \Ref{linexp}), 
the quadratic one maintains the right asymptotic behaviour. This important 
property was obtained for the vacuum theory in \cite{Krasnov2}, and here 
we extend it to the coupling to electromagnetism by a perturbative analysis. 
Below in Section \Ref{SecRN} we show it also using the non-perturbative 
numerical solution. Before studying numerically the system, let us discuss 
the vacuum case $Q=0$, that can be solved analytically as found 
in \cite{Krasnov2}.

%%%%%%%
\section{Vacuum black holes revisited} \label{SecVacuum}
%%%%%%%
In \cite{Krasnov2}, Krasnov and Shtanov solved the system \Ref{MaxQ} for 
$Q=0$ and found an analogue of the Schwarzschild black hole. Here we review 
their results and extend them to include the locations of the event horizon. 
The knowledge of the analytic solution allows us to test our numerical 
integration code, which we use for the plots of this Section and the next 
Section on the charged case. 

\medskip 
The Krasnov-Shtanov solution is given by
\equ
x(r) = e^{-\f12 A \beta} \,, 
\nequ
with $\beta(r)$ implicitly defined by
\equ\label{betaK}
r(\beta) = \left(\f{M}{\beta}\right)^{\f13} e^{\f{A}6 \beta} \,,
\nequ
$M$ being an integration constant, corresponding to $M$ 
in eq.~(\ref{betapert}).\footnote{This solution can be expressed in terms of the Lambert $W$ function (see e.g. \cite{Lambert}),
$$
\beta(r) = -\f{2}{A} \, W\left(-\f{AM}{2r^3}\right).
$$
We thank Robert Mann for pointing this out to us.} 
From \Ref{MaxQ3} with $Q=0$ we can immediately extract the non-trivial components of the metric,
\eqa\label{g00K}
g_{00}(r) &=& \f1{c^2} \, f_*^2 
           = \f{2(1+A \beta)}{2-A \beta} \, e^{-{A} \beta} 
             \Big[
                  1- (2\beta - \f12 A \beta^2 - \Phi_0) r^2 
             \Big] \,, \\ \label{g11K} 
g_{11}(r) &=& -\f1{c^2} \, g_*^2 
           = -\f{2(1+A \beta)}{2-A \beta} \, 
              \Big[
                   1- (2\beta - \f12 A \beta^2 - \Phi_0) r^2   
              \Big]^{-1} \,.  
\neqa
One can easily check that setting $A=0$ gives back the standard 
Schwarzschild-de Sitter metric representing a black hole of mass $M$. 
\footnote{ %%% 
Note that here and in the following, by $g_{00}$ and $g_{11}$ we mean 
the $(t,t)$- and $(r,r)$-coordinate components of the metric;  
they are not the components with respect to frames, 
$\{e^I \}$, or $\{l^I\}$, introduced in Section~\ref{SecSph}. 
} %%% 

\medskip 
At values of $r$ where $1-(2\beta(r) - \f12 A \beta^2(r) - \Phi_0) r^2=0$,  
we have a coordinate singularity that can be removed via a change of 
coordinates. If $c^2>0$ outside, then the coordinate singularity may be 
interpreted as the event horizon. The metric becomes degenerate also at 
the points such that $\pb = A\beta=2,-1$. 
These are non-metric singularities where $c^2 \rightarrow 0$ (type~(1)) 
or $c^2 \rightarrow \infty$ (type~(2)), and where the relation between 
the fundamental field $B$ and the metric breaks down. 
To see whether and when these singularities occur, note first 
that the solution for $\beta(r)$ obtained inverting eq.~\Ref{betaK} is always 
positive for $r>0$. This means that $A\beta=2$ can only 
occur if $A>0$, and the one at $A\beta=-1$ if $A<0$. 
Recalling that $\Lambda=-3\Phi_0$, a positive (resp. negative)
$A$ has a coherent sign with a negative (resp. positive) cosmological constant.

\medskip 
Note that $\Phi_0$ enters the equations in the same unmodified way that the 
cosmological constant does in general relativity. In the following we will 
restrict attention to $\Phi_0=0$ and focus on the effect of the modification 
introduced by a non-constant $\pb$. 

\medskip 
What is the nature of the non-metric singularity? At these locations, 
the metric becomes singular or divergent, see eqs.~(\ref{g00K}-\ref{g11K}). 
However it was shown in \cite{Krasnov2} that these are not physical 
singularities for the theory: a change of coordinates exists such that all 
the fundamental fields $B$ and $\om$ are finite. To see supporting evidence 
that the singularity is not physical, we can also look at scalar invariants. 
One simple scalar invariant can be constructed out of the field 
equations \Ref{EqF2} as follows: 
first, contract the equations with the wedge product of $B^i$. 
This gives a scalar density. Second, divide by the volume element provided by
$V=\f13 B^i\w B_i$. 
In the standard general relativity this procedure would give the Ricci scalar.
In the modified theory, it defines a scalar quantity 
${\cal S}\equiv B^i\w F_i / V$. 
Taking $\Phi_0=0$ for simplicity, the equations \Ref{EqF2} give after some 
algebra 
\equ\label{scalar}
{\cal S} = \f92 A \beta^2 \,. 
\nequ
This shows explicitly the finiteness of this scalar quantity at the non-metric 
singularities of both type (1) $A\beta=2$ and (2) $A\beta=-1$, 
and justifies, in part, the claim that these non-metric singularities are 
not physical singularities of the theory. 
What is happening is not that the theory is breaking down, but rather at the 
locations of these non-metric singularities, there doesn't exist any metric 
with respect to which the 2-forms, eq.~\Ref{Bsph}, is self-dual. 

\begin{figure}[h!]\begin{center}
\includegraphics[width=4cm]{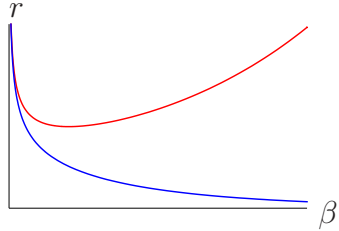}
\begin{picture}(0,0)(0,0)\put(0,-5){\large$\beta$}\put(-117,73){\large$r$}\end{picture}
\caption{\small{Paradigmatic plots of $r(\beta)$ from \Ref{betaK}, 
for $A>0$ (red) and $A<0$ (blue). In the first case a minimal possible 
radius arises. % and a range of invertibility is $A \beta \in (0,2)$. 
In the second case the function is always invertible. %, and $A \beta \in (-\infty,0)$.
}\label{Figbeta} }
\end{center}
\end{figure}

\medskip 
The two cases of positive and negative $A$ are qualitatively different 
(see Fig.~\ref{Figbeta}), and it is convenient to treat them separately. 

%%%%%%%%%%%
\subsection{Positive $A$ branch}
%%%%%%%%%%%

\begin{figure}[h!]\begin{center}
\begin{picture}(0,0)(0,0)\put(272,8){$A$}\put(90,173){$r$}
\put(115,130){$\downarrow$}\put(115,140){$\rH$}\put(135,100){$\longleftarrow \rnm$}\end{picture}
\includegraphics[width=9cm]{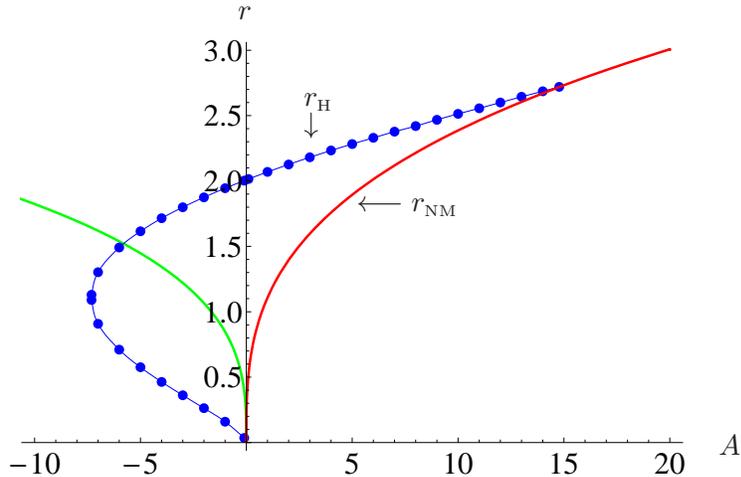} 
\caption{\label{FigrUnico} 
\small{Location of the horizons and non-metric singularities as functions of 
$A$, for $M=1$. The red continuous curve is the non-metric singularity 
$\rnm$ for $A>0$ and the green continuous curve for $A<0$. 
The blue interpolated dots are numerically found zeroes of $f_*^2$ 
representing the radius $\rH$ of the event horizon. 
For negative $A$, we have two horizons. 
The values of $A$ which undress the singularity compute analytically to
$A=2e^2\simeq 14.78$ and $A=-125/(8e) \simeq -5.75$. 
Analytically we can compute that the two zeroes of $f_*^2$ present for 
negative $A$ merge at $A=-54 e^{-2} \simeq -7.31$. Yet notice that the zeroes 
numerically found do not match exactly at this value: 
this shows explicitly the limitation of the numerical integration. The error
can be estimated to be $\sim 5\%$.
}}
\end{center}
\end{figure}

\begin{figure}[h!]
\begin{center}
\begin{picture}(0,0)(0,0)\put(218,38){$r$}\put(23,118){$f_*^2$}\end{picture}
\begin{picture}(0,0)(0,0)\put(-50,50){\framebox{$A>0$}}\end{picture}
\includegraphics[width=7cm]{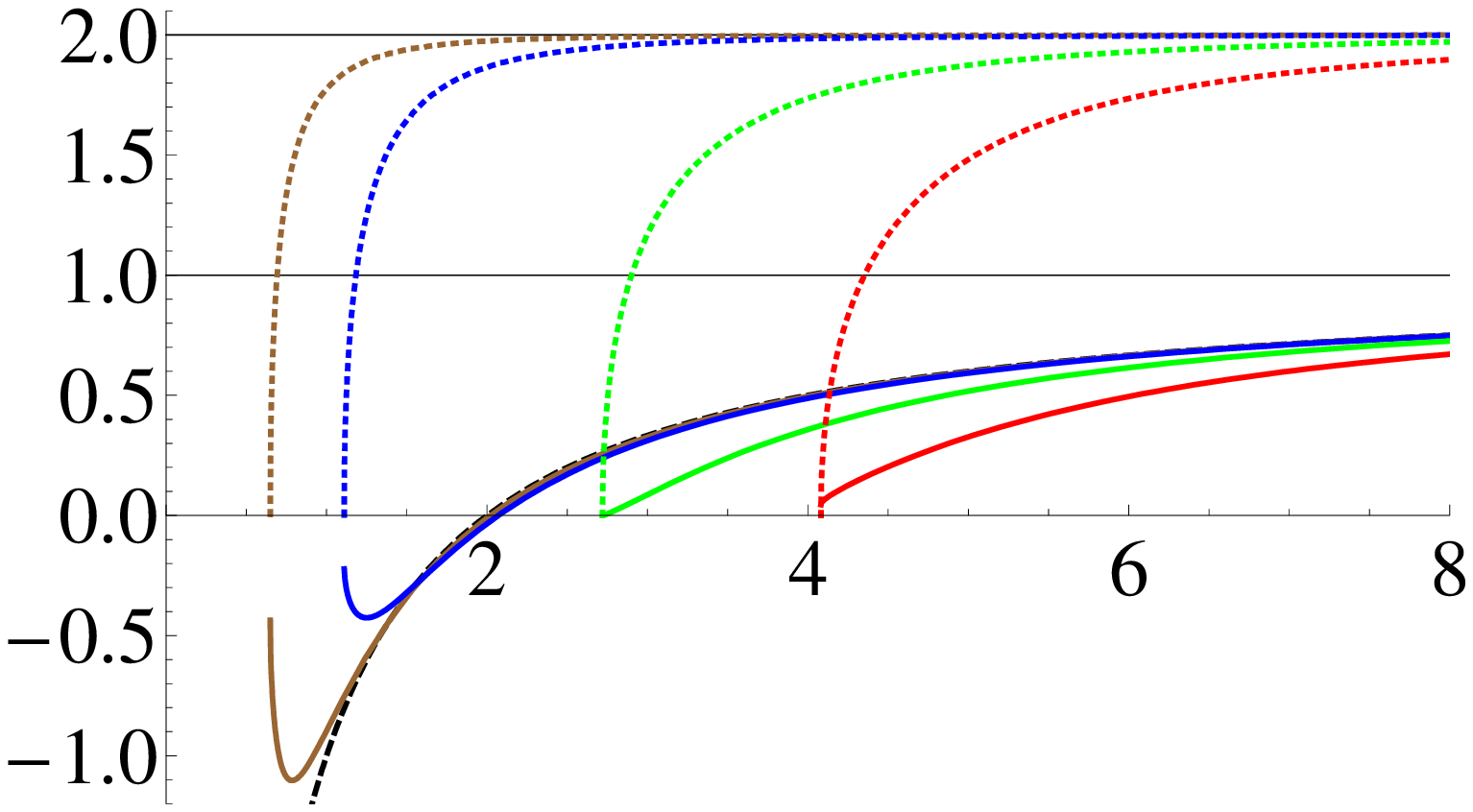}
\caption{\label{FigfsA} 
\small{Plots of $f_*^2$ (continuous) and $2-A\beta$ (dotted)
 for different values of $A>0$, with $M=1$. From left to right, $A=0.2$ (brown),
 $A=1$ (blue), the limiting case $A=2e^2$ (green), and $A=50$ (red).
 The dashed black line is the standard general relativity result, drawn here for comparison
 of the asymptotic behaviour.
 The zeroes of the dashed lines are the locations of the non-metric singularities $A\beta=2$, where $c^2=0$
 and this branch of the solution interrupts.
}}
\end{center}
\end{figure}

When $A$ is positive we see from Fig. \ref{Figbeta} that $r(\beta)$ 
in eq.~\Ref{betaK} has a minimum. Consequently, there is a minimal value of 
$r$ that can be reached. Imposing the derivative of eq.~\Ref{betaK} to vanish, 
this minimum is found at $\rnm = (e A M / 2)^{1/3}$. We call this point 
`$\rnm$' because it is precisely where $A\beta=2$, thus at this point 
there is a non-metric singularity of type (1). $\beta$ can be inverted either 
in the domain $r\in (\rnm, \infty)$, or $r\in (0,\rnm)$. 
The first one has the standard asymptotic behaviour 
$\beta \sim r^{-3}\; (r\mapsto \infty)$ of general relativity, 
and we look at this case first. In this branch we have $A \beta \in (0, 2)$, 
thus $c^2$ is always positive; consequently, the outmost zero of $f_*^2$ 
corresponds to the event horizon, and changes in the signature of the metric 
come from changes in the sign of $f_*^2$ only. 
It is natural to ask whether the singularity at $\rnm$ lies within 
the event horizon, or it is naked. 
The horizon is located at $f_*^2(\rH)=0$, but this equation cannot be 
solved analytically. 
To visualise the situation, we find numerically the zeroes $\rH(A)$ of 
$f_*^2$ and compared them with $\rnm(A)$. 
See Figure \ref{FigrUnico}. 
To further illustrate the structure of the spacetime metric,
we plot $\beta$ and $f_*^2$ for various values of $A$ in Fig. \ref{FigfsA}.
The complete behaviour of the metric can be easily inferred from these plots 
and eqs.~(\ref{g00K}) and (\ref{g11K}). 

\medskip 

\medskip 
We see from Fig. \ref{FigrUnico} that both $\rH(A)$ and $\rnm(A)$ increase 
with $A$. The radius of the non-metric singularity grows in $A$ faster than 
the horizon radius, 
and there is a value of $A$ for which the radius of the singularity reaches 
the horizon radius. This value can be computed by requiring $f_*^2(A\beta=2)$ 
to be non-positive (see Fig. \ref{FigfsA}), and one finds $A \leq 2 e^2 M^2$. 
This means that for fixed $A$ there is a minimal mass, 
$M_0 = \sqrt{{A}/(2e^2)}$ for which the black hole exists: 
for smaller values of the mass, the non-metric singularity is naked. 

\medskip 
At the non-metric singularity both $g_{00}$ and $g_{11}$ 
are diverging, but this is not a physical singularity as we have discussed 
above. Hence the theory itself is not breaking down and the solution can be 
continued to describe a ``spacetime'' beyond this point. 
From the point of view of the metric, continuing the solution means 
using the other branch of the inverse of $r(\beta)$ in eq.~\Ref{betaK}. 
This is the branch where $A\beta\in(2,\infty)$ and we move from 
$r=r_{\rm NM}$ to $r=\infty$ as we recede from the non-metric 
singularity. Since $c^2$ is always negative there, it follows from 
eqs.~\Ref{g00K} and \Ref{g11K} that the behaviour of the metric inside 
the non-metric singularity is similar to the case of Kantowski-Sachs 
cosmological spacetime, with the singularity at $r \rightarrow \infty$ 
(and thus $\beta \rightarrow \infty$, $g_{00} \rightarrow 0^-$, 
$g_{11} \rightarrow 0^+$) being a true curvature singularity.    
We do not report here plots of the solution in this branch.
The overall picture can be summarised in the Penrose diagram 
already present in \cite{Krasnov2} (e.g., for the region of Kantowski-Sachs 
spacetime, see VI and VIII in Fig. 1 of \cite{Krasnov2}), 
where further discussion of the global structure is also given.

%%%%%%%% 
\subsection{Negative $A$ branch} 
%%%%%%%% 
When $A$ is negative, the function $r(\beta)$ in eq.~\Ref{betaK} is invertible 
for any $r>0$. Since $A\beta\in(-\infty,0)$, the $A\beta=2$ singularity cannot 
occur, but rather the $A\beta=-1$ one. At large radii, $\beta$ approaches zero 
with the standard $r^{-3}$ behaviour. As we decrease $r$, we reach the value 
$\rnm$ at which $A\beta=-1$ and the type~(2) non-metric singularity occurs. 
The location is easily determined plugging $A\beta=-1$ in \Ref{betaK}: 
we find $\rnm = (M |A|)^{1/3} e^{-1/6}$. There, $c^2=0$, and both $g_{00}$ 
and $g_{11}$ vanish. As already seen, this is not a physical singularity of 
the theory. The solution continues after this singularity (recall $\beta'$ 
never vanishes, so eq.~\Ref{betaK} can be inverted for all $r>0$), 
with $c^2<0$ and the metric signature is switched to $-+--$ as $\beta$ 
becomes large.

\begin{figure}[h!]\begin{center}
\bigskip
\begin{picture}(0,0)(0,0)\put(213,50){$r$}\put(20,118){$f_*^2$}
\put(-50,50){\framebox{$A<0$}}\end{picture}
\includegraphics[width=7cm]{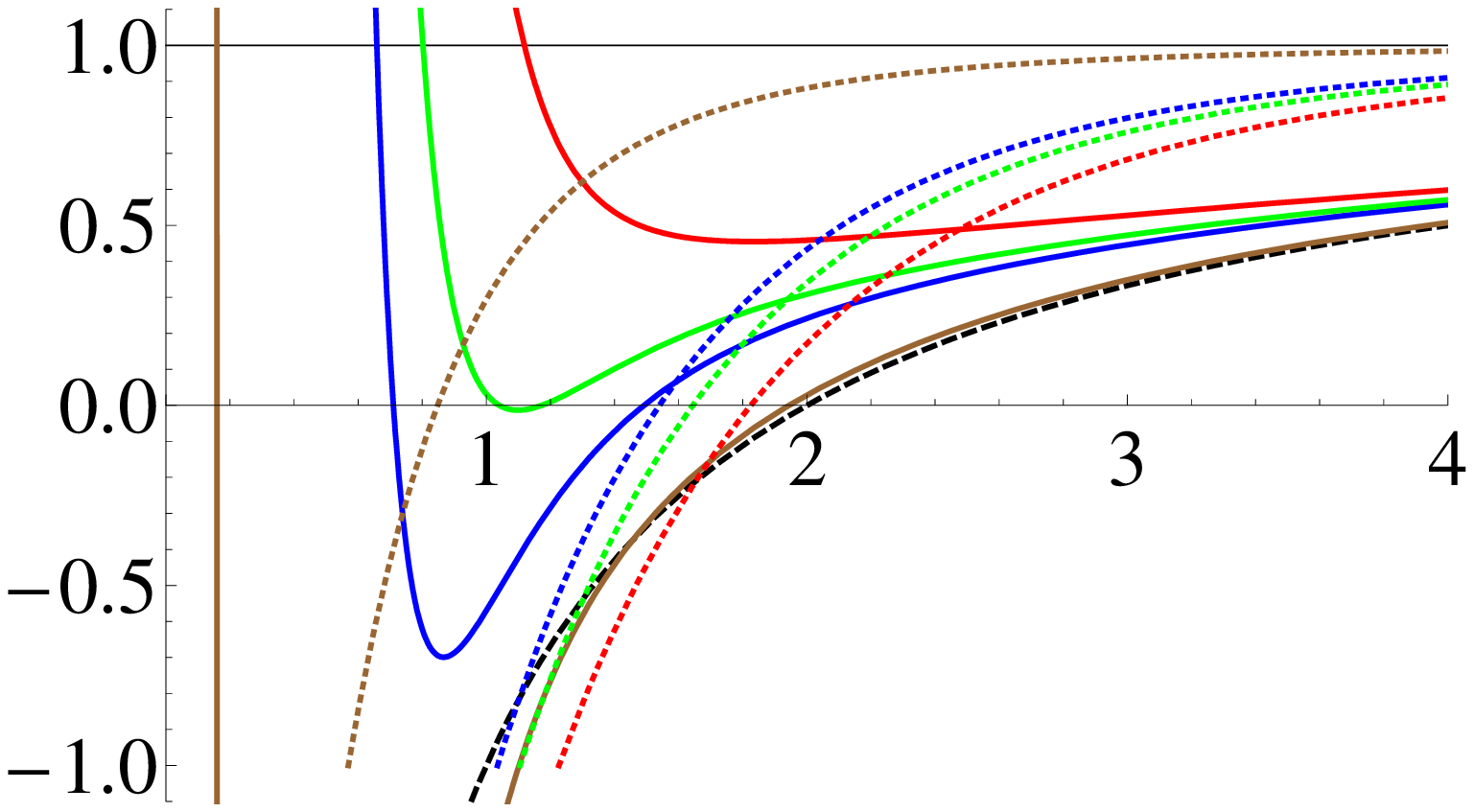}
\caption{\label{FigAN} 
\small{Plots of $f_*^2$ (continuous) and $1+A\beta$ (dotted) for different 
values of $A<0$, with $M=1$. From left to right, $A=-1$ (brown), $A=-6$ 
(blue), the limiting case $A=-54 e^{-2}$ (green), and $A=-10$ (red). 
Notice that the green curve has not precisely a single horizon, 
but rather two: This shows the same numerical limitation exposed in 
Fig.~\ref{FigrUnico}. The dashed black line is the standard general relativity 
result, showing the matching asymptotic behaviour. The zeroes of the dashed 
lines are the locations of the non-metric singularities of type~(1), 
$A\beta=2$. There, $c^2$ diverges, $f_*^2$ is finite, and thus the metric 
components $g_{00}$ and $g_{11}$ both vanish.
}}
\end{center}
\end{figure}

\medskip 
Again, we can numerically find the zeroes of $f_*^2$ to see whether the 
type~(2) non-metric singularity is hidden or not by the event horizon, see 
Fig.~\ref{FigrUnico} for $A<0$. 
In Fig.~\ref{FigAN}, we plot $f_*^2$ and $\beta$. 
The plots show that $f_*^2$ approaches the Schwarzschild solution at large 
distances, and hence we can naturally expect that all solutions are 
asymptotically flat. 

The plots also show the near horizon structure, the relative positions of 
the horizon, and the non-metric singularity, as zeroes of $f_*^2$ and 
$1+A\beta$ (see also Fig. \ref{FigrUnico}). 
Notice the presence of \emph{two} zeroes of $f_*^2$. For $A$ small enough, 
the non-metric singularity lies inside the event horizon given by the 
outmost zero of $f_*^2$, i.e., at the outmost zero of $f_*^2$, $1+A\beta>0$. 
We find that the event horizon is at a smaller radius than the Schwarzschild 
one. As we increase $A$, the radius of the event horizon shrinks,  
whereas that of the non-metric singularity grows. The critical value of $A$ 
at which the singularity becomes naked can be computed by requiring 
$f_*^2(\rnm)\leq 0$, and one gets $|A| \leq \left(\f52\right)^3 \f{M^2}{e}$. 
Therefore we have a situation similar to the positive $A$ branch with 
the minimal black hole, having this time the mass 
$M_0 = \left(\f25\right)^3/2 \sqrt{e |A|}$.  
For larger values of $|A|$ the non-metric singularity $A\beta = -1$ is 
exposed, and the two zeroes of $f_*^2$ keep moving towards each other. 
The two zeroes merge in a similar manner that two Killing horizons 
of the Reissner-Nordstrom metric merge and form a single degenerate horizon 
when the extremal limit $|Q| \rightarrow M$ is taken. However, for the present 
case, both the zeros of $f_*^2$ occur in the region where $1+A\beta <0$: 
both the ``horizons'' defined as the zeroes of $f_*^2$ are located inside 
the non-metric singularity. 
%In other words, the non-metric singularity is exposed as a naked singularity. 
The value at which the two zeros of $f_*^2$ merge and disappear can be 
computed by requiring the minimum of $f_*^2$ to be non-negative. We find 
$A\leq -54M^2/e^2 \simeq -7.31 M^2$ (green line). 
In \cite{Krasnov2} this relation was interpreted to give, for fixed $A$, 
the minimal value of the mass for the black hole to exist. 
However we have seen above that well before 
(at $|A|= \left(\f52 \right)^3 \f{M^2}{e}\simeq -5.7 M^2$), 
the black hole has already ceased to exist because the non-metric singularity 
at $\pb=-1$ is naked. After the zeroes of $f_*^2$ disappear even the true 
singularity at $r=0$ is exposed.

\begin{figure}[h]\begin{center}
\includegraphics[width=6cm]{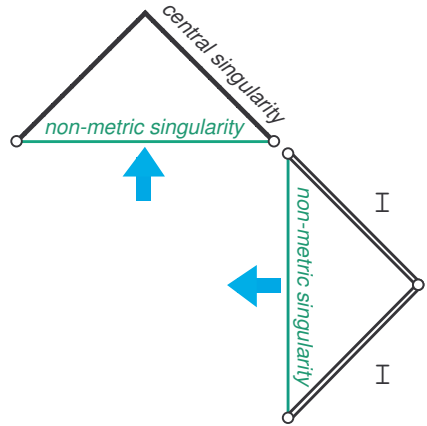} \hspace{1cm} 
\caption{\small{
Conformal diagram for the vacuum ($Q=0=\Lambda$) solution with a naked 
non-metric singularity (green line), originally shown in \cite{Krasnov2}. 
The double line, ${\cal I}$, denotes null infinity. 
The flow of time is vertical, different from the original diagram given 
in \cite{Krasnov2}, and the two green lines should be identified.  
It is seen that the naked non-metric singularity (vertical green line) 
corresponds to a ``Cauchy surface'' (horizontal green line) for the interior 
region, and accordingly the central singularity (thick dashed line) is 
achronal (non-timelike).    
}\label{fig:A1naked} } 
\end{center}
\end{figure}

\medskip
Summarising with the help of Fig. \ref{FigrUnico}, we see that in both 
branches we have at fixed $M$ a maximal value of $|A|$ above which the black 
hole ceases to exist. This means that for a given $A$ there is always a 
minimal mass that a black hole can have, $M_0 = \sqrt{{A}/(2e^2)}$ for 
positive $A$ and $M_0= \left(\f25\right)^{3/2} \sqrt{e |A|}$ for negative $A$. 
In the range of $A$ for which the horizon exists, its location at fixed $M$ is 
an increasing function of $A$. It is smaller than the Schwarzschild value 
$\rH=2M$ for $A<0$ and larger for $A>0$. 
The spacetime structure in the case of a naked non-metric singularity with no horizons present
is shown in the conformal diagram in Figure \ref{fig:A1naked}.

%-----------------------------------------------------
%%%%%%
%-----------------------------------------------------

\newpage

%%%%%%%%
\section{Charged black holes}\label{SecRN}  
%%%%%%%% 
We now consider the full system of equations \Ref{MaxQ} in the presence of 
an electric charge. The numerical studies that we present below show that 
the metric components approach those corresponding to the Reissner-Nordstr\o m 
metric at large distances, and that $\beta$ goes to zero, thus $c^2>0$ and 
the asymptotic signature is $+---$. This confirms the asymptotic flatness of 
the perturbative solution presented earlier (see eq.~\Ref{betapert}). 
Thanks to this asymptotic behaviour, which conforms to 
the conditions \Ref{condi:flat:c}, \Ref{condi:flat:fg}, \Ref{condi:flat:beta}, 
the parameters $M$ and $Q$ can be interpreted as the ADM mass and the electric 
charge in the standard sense discussed in Section~\ref{sect:AF}. 
\footnote{ %%% 
The analytic perturbative solution \Ref{betapert} 
allows us to identify $M$ and $Q$ as the ADM total mass and charge. However 
for practical purposes in the numerical code one has to assign the boundary 
conditions at a finite value of the radius, say $r_0$. The larger is $r_0$, 
the closer the boundary values $M$ and $Q$ are to the ADM charges; 
but also the less precise are the numerical solutions near the horizon.
Conversely, a small $r_0$ increases the precision of the 
solution near the horizon, but limitates the interpretation of $M$ and $Q$. 
After investigating the stability of our code (by comparing it with the 
analytical cases known at $Q=0$, see left panel of Fig.\ref{FigAN}) 
we took $r_0=50$. 
} %%% 

\medskip 
Decreasing $r$ from the asymptotic regime, the curvature increases and we 
start seeing departures from the Reissner-Nordstr\o m solution in general 
relativity. In particular, 
\begin{itemize} 
\item The number and the location of the horizons are changed; 
\item There appear again non-metric singularities at $A\beta=2$ (of type~(1)), 
or $A\beta=-1$ (of type~(2)). 
\end{itemize}
Concerning the non-metric singularities, they now occur at locations 
$\rnm(A,M,Q)$ depending also on $Q$. 
At these points the metric given by \Ref{g00} is degenerate or diverging, 
but as seen in the uncharged case, this does not mean that the singularities 
are harmful to the theory itself. In this perspective, we point 
out that the scalar $\cal S$ constructed in \Ref{scalar} is unchanged by 
the presence of an electromagnetic field, thanks to the property 
$B^i \w \tau_i = 0$. This is analogue to what happens in general relativity, 
where the presence of an electric charge does not affect the Ricci scalar.  
Therefore $\cal S$ is again finite at the non-metric singularities. 
In view of this, 
although we have not fully studied the behaviour of other curvature components 
at the singularities, we anticipate that a similar change of coordinates 
as the one made in \cite{Krasnov2} would show finiteness of the fundamental 
fields of the theory.  

\medskip 
As in the uncharged case, the solutions in radial coordinate $r$ interrupt at 
$A\beta=2$, where $\beta'$ diverges (see eq.~\Ref{MaxQ2}), whereas they 
continue after $A\beta=-1$. So again the range of our plots is 
$r\in(0,\infty)$ if no type (1) singularity occurs, whereas it is 
$(\rnm(A\beta=2),\infty)$ if it does. However, unlike the uncharged case, 
both types of non-metric singularities can occur regardless of the sign of 
$A$, since now the sign of $\beta$ is not fixed (see e.g., the perturbative 
solution \Ref{betapert}). 

\begin{figure}[h]\begin{center}
\includegraphics[width=7cm]{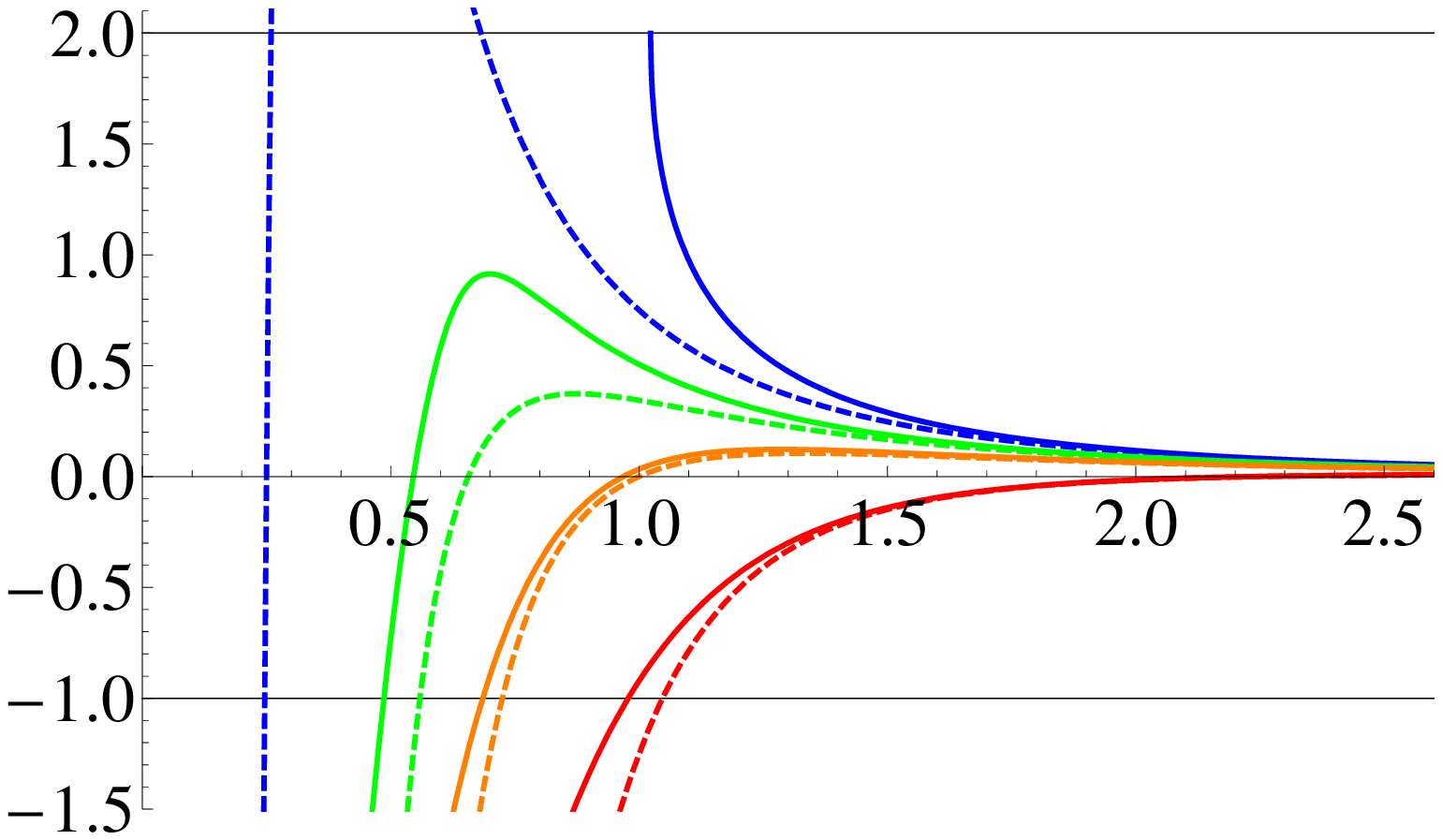} \hspace{1cm} 
\includegraphics[width=7cm]{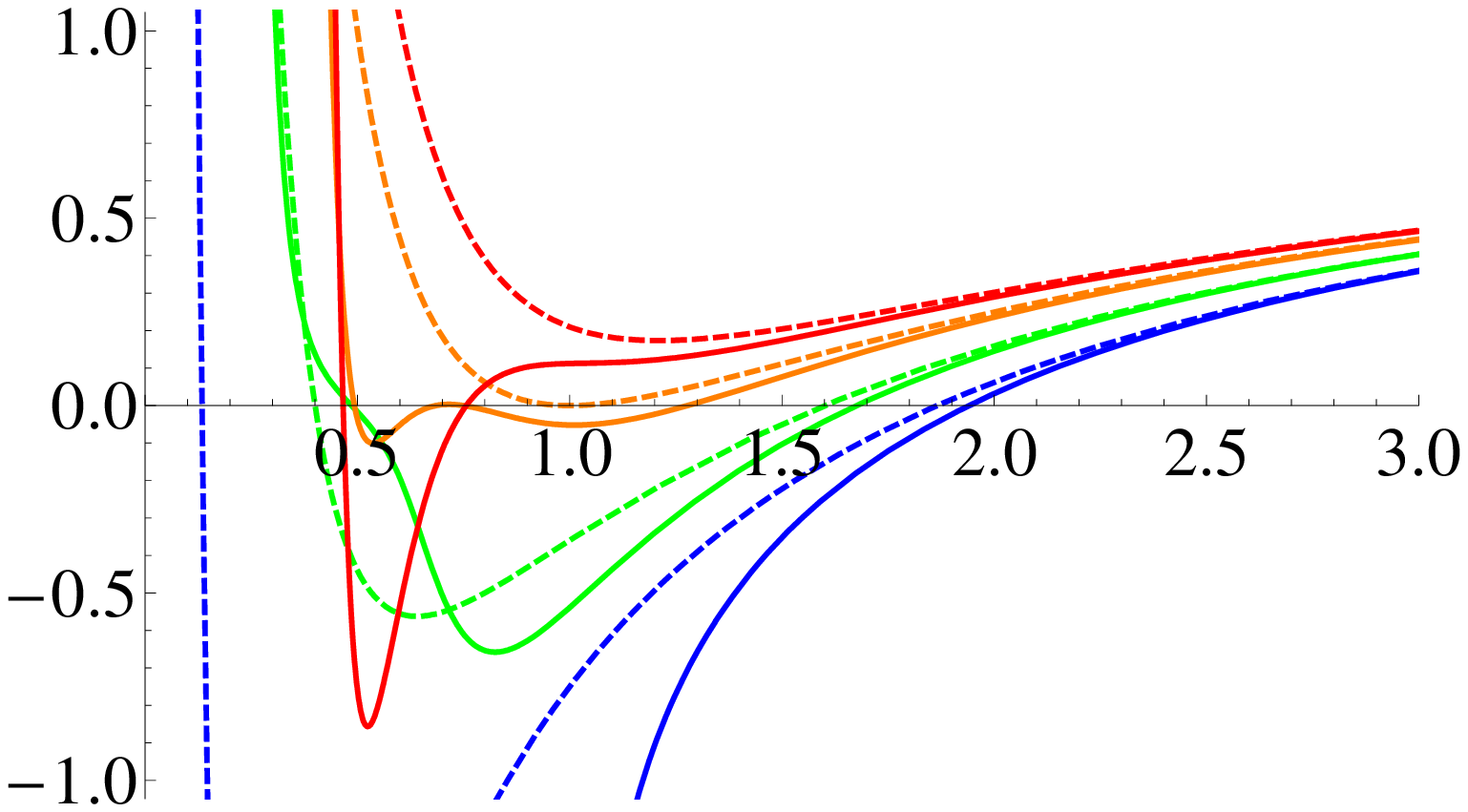}
\begin{picture}(0,0)(0,0)\put(0,50){\large$r$}\put(-190,118){\large$g_{00}$}\end{picture}
\begin{picture}(0,0)(240,0)\put(0,48){\large$r$}\put(-190,118){\large$A\beta$}\end{picture}
\caption{\small{
The numerical solutions for $M=1$ and $A=1$ of the modified theory 
(solid lines) plotted against the Reissner-Nordstr\o m solution of general 
relativity (dashed lines) for $Q=0.5$ (blue), $Q=0.81$ (green), $Q=1$ (orange) 
and $Q=1.5$ (red). Note the matching asymptotic behaviour. $g_{00}$ diverges 
at minus infinity at the radius where the type~(1) $A\beta=2$ singularity 
occurs (blue line), whereas diverges at plus infinity at $r=0$, as the 
Reissner-Nordstr\o m solution, when this type~(1) singularity is avoided. 
In the latter case, extra zeroes may appear in $g_{00}$ due to crossings of 
the type~(2) $A\beta=-1$ singularity. The value $Q=0.81$ estimates the turning 
point from one regime to the other. 
}\label{Figg00A1} }
\end{center}
\end{figure}

\medskip 
To get a first feeling of departures from the general relativity case, 
in Figure \ref{Figg00A1}, we plot the numerical solutions of $g_{00}$ 
for $A=M=1$ and various values of the charge, and compare them with 
the Reissner-Nordstr\o m solution in general relativity. 
These plots show the matching asymptotic behaviour anticipated above. The 
behaviour of $g_{00}$ near the horizon, on the other hand, becomes rather 
different in the modified theory. Note that zeroes of $g_{00}$ are not 
necessarily horizons, as they can also correspond to a non-metric singularity 
of type~(2). 

\begin{figure}[h]\begin{center}
\includegraphics[width=7cm]{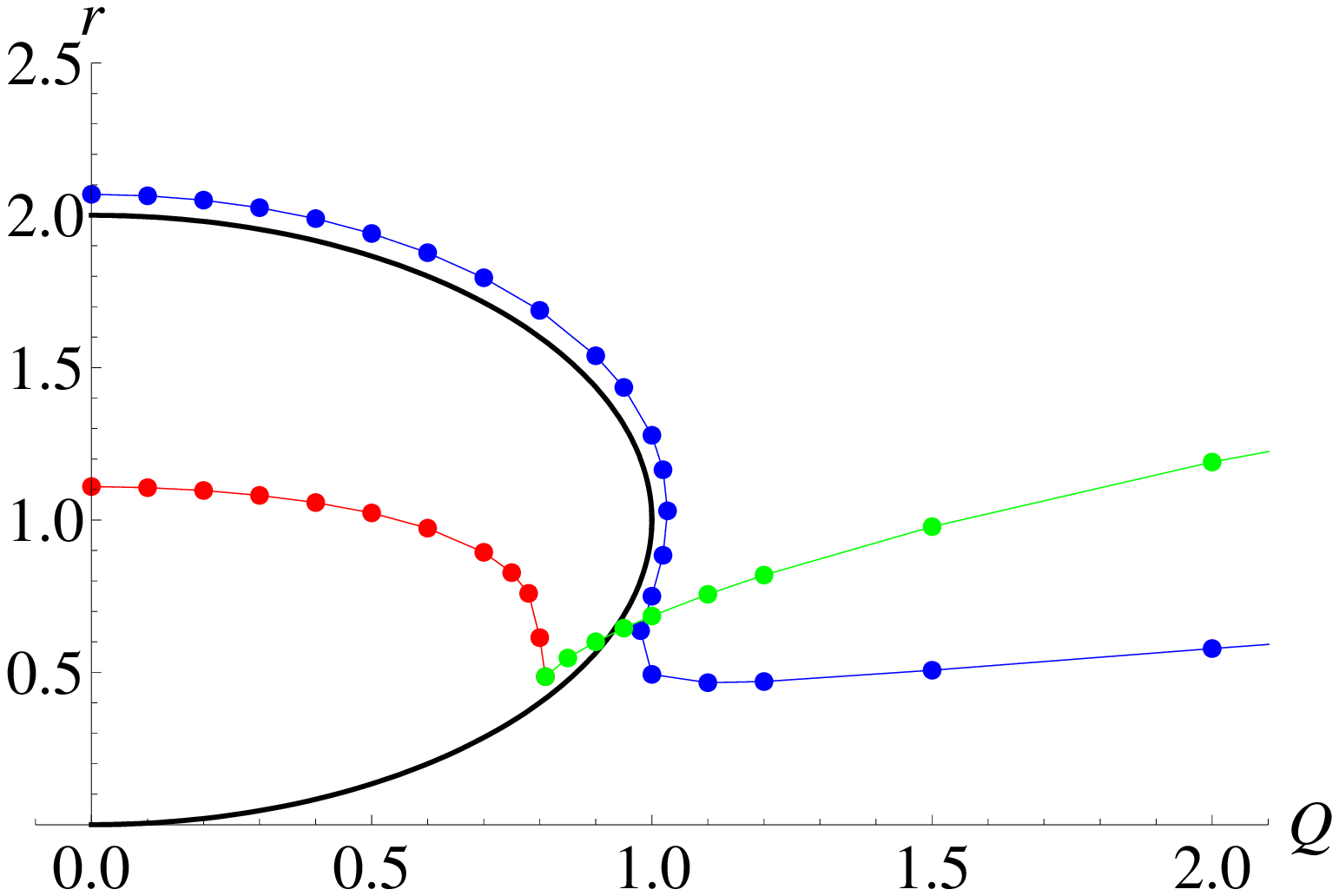} \hspace{1cm} 
\includegraphics[width=7cm]{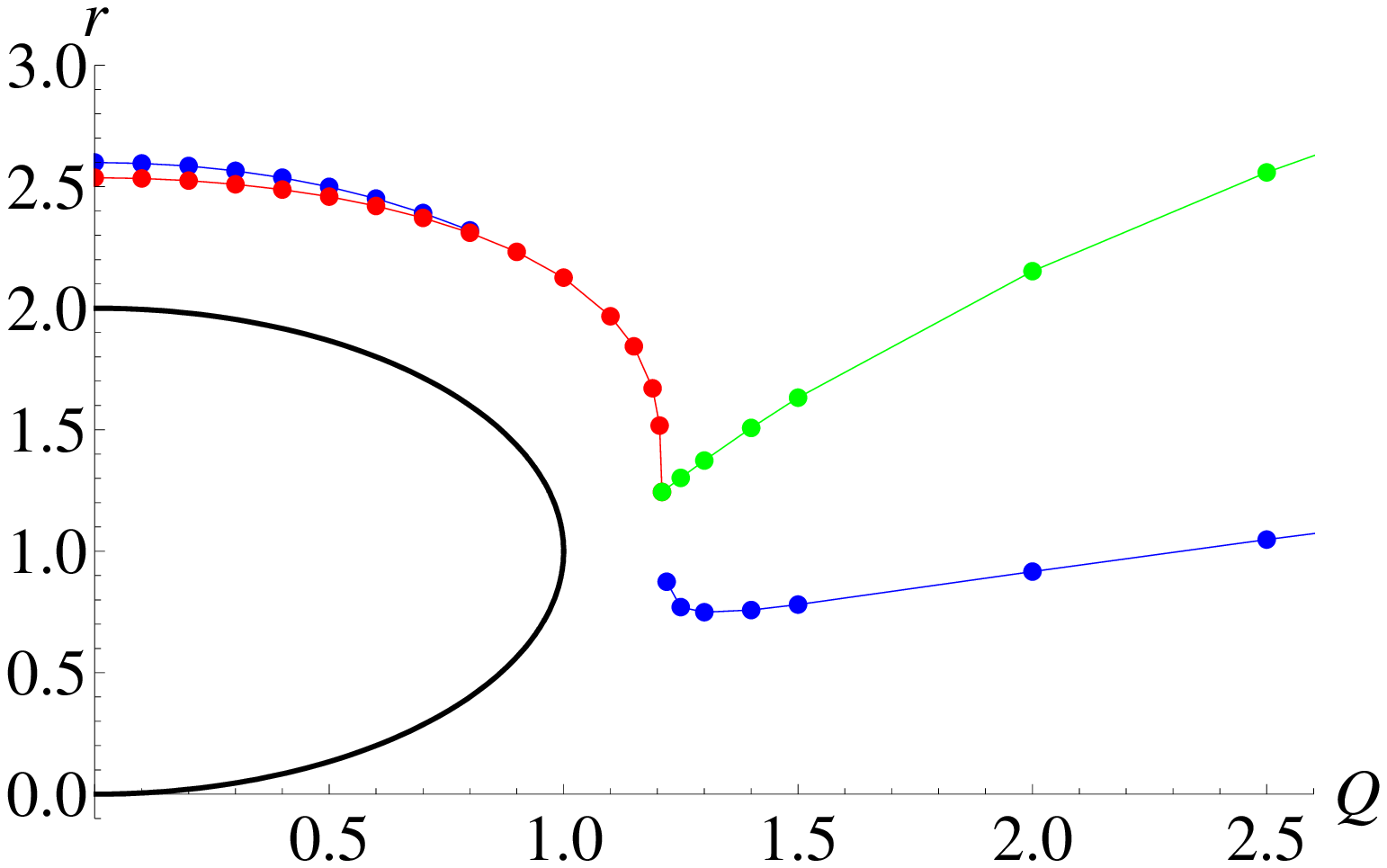} 
\begin{picture}(0,0)(0,0)\put(-340,125){\framebox{$A=1$}}\put(-110,125){\framebox{$A=12$}}\end{picture}
\begin{picture}(0,0)(0,150)\put(-340,130){\framebox{$A=-1$}}\put(-110,130){\framebox{$A=-7$}}\end{picture}
\includegraphics[width=7cm]{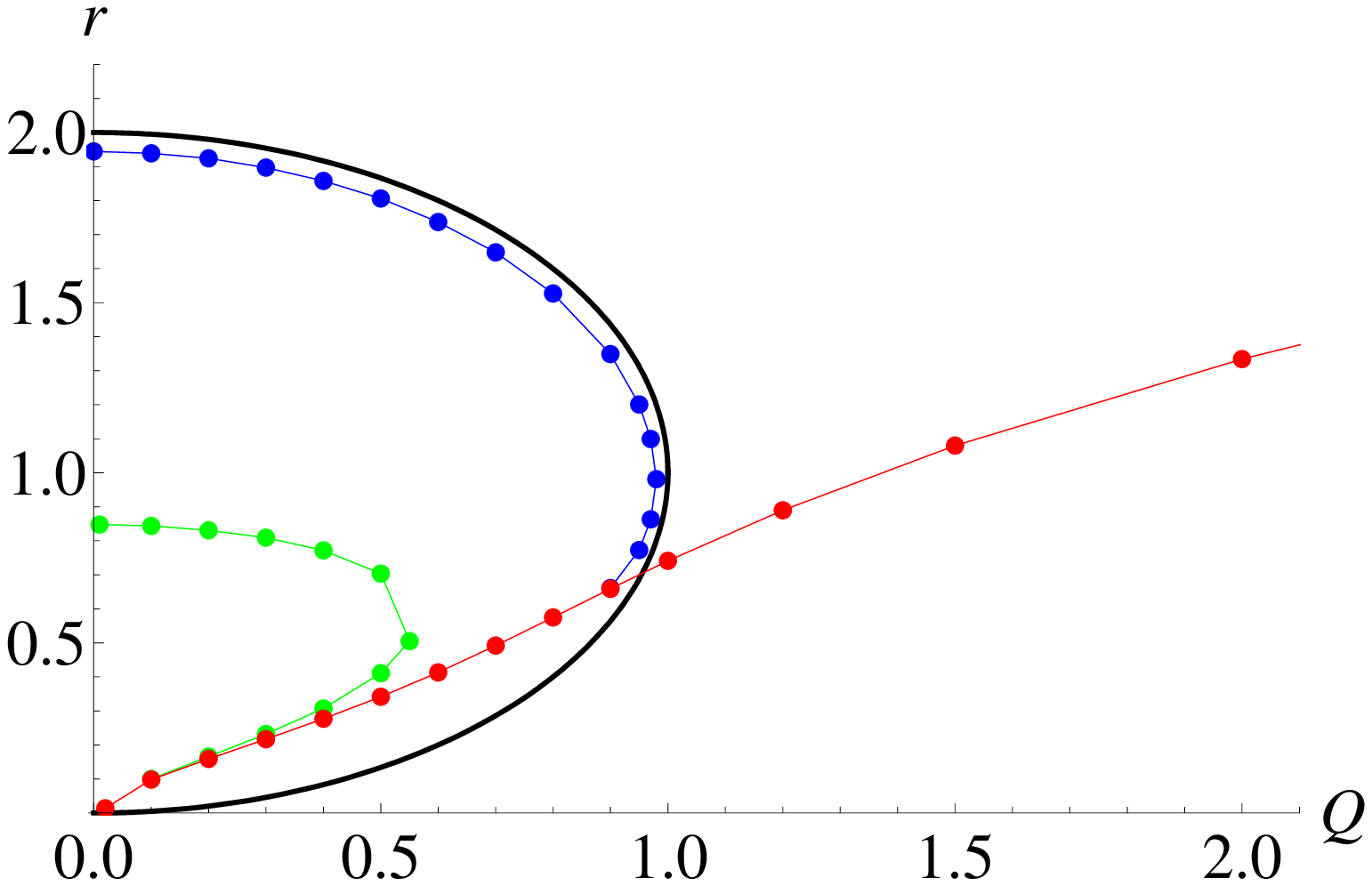} \hspace{1cm} 
\includegraphics[width=7cm]{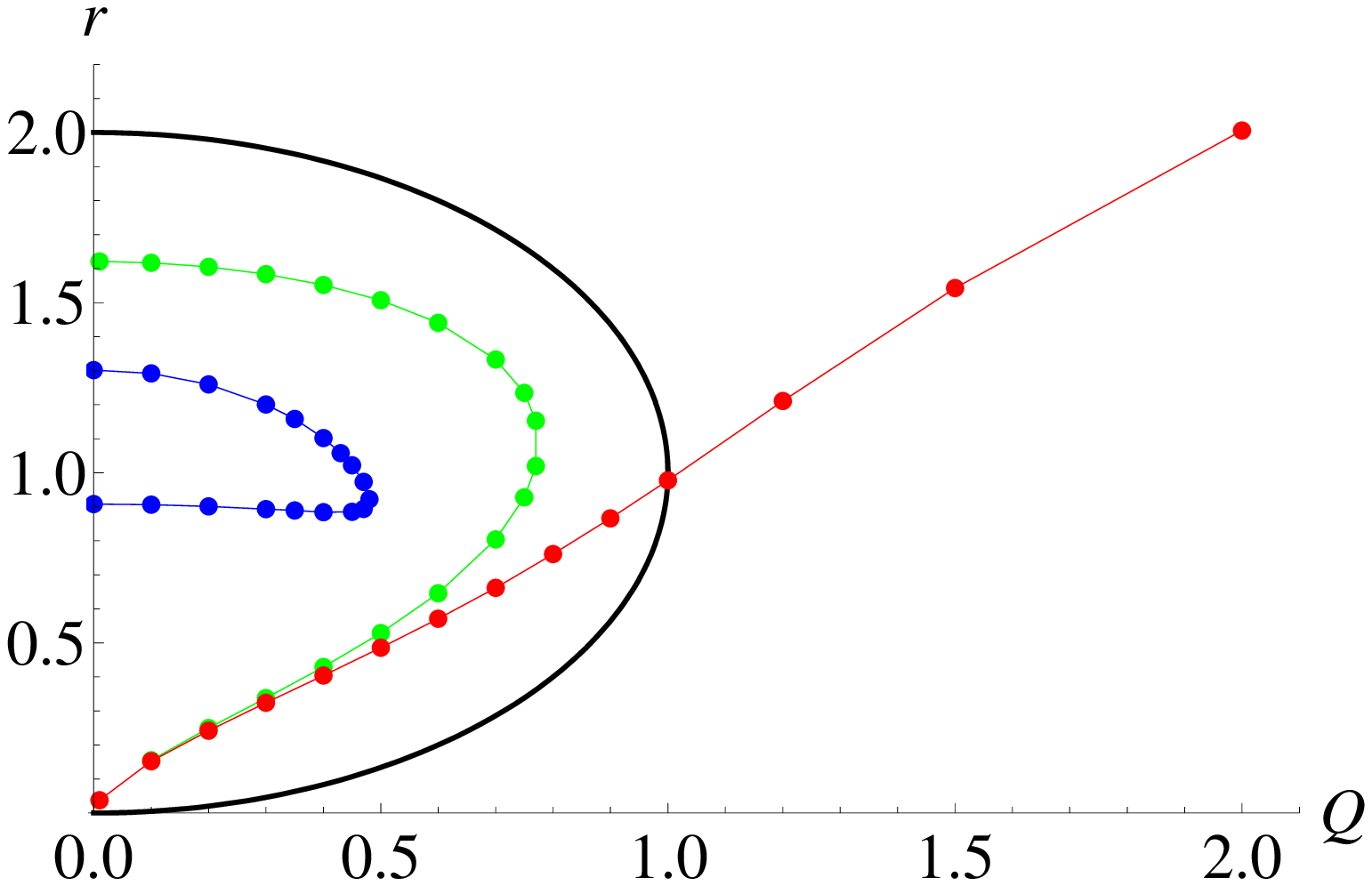}
\caption{\label{FigrApos} 
\small{Locations of the event horizons, interpolated blue dots, and non-metric 
singularities of type~(1) (red dots), and type~(2) (green dots), at $M=1$. 
For comparison, the horizons of the Reissner-Nordstr\o m solution are also 
plotted as black lines. 
}}
\end{center}\end{figure}

\medskip 
As we vary $A$, the departures from general relativity change significantly,
and we have to face rich fauna of solutions. 
To give a useful overview of the situation, we plot in Figure \ref{FigrApos} 
the locations of horizons and non-metric singularities as functions of 
$Q$, for a few significant values of $A$. We also illustrate 
the global spacetime structures for a few specific cases in 
Figures \ref{fig:A1nakedhorizon}, \ref{fig:A1horizon}, 
and \ref{fig:A3horizons}. 

\medskip 
The initial parameter space in which the numerical integrations are performed  
is four-dimensional: $(\Phi_0, A, M, Q)$. However $\Phi_0$ and $M$ affect 
the solutions in a trivial way: 
a non-vanishing cosmological constant $\Phi_0$ can be straightforwardly added 
using \Ref{MaxQ3}, and when changing the boundary condition $M$, we simply 
reproduce the same features at appropriately rescaled $A$ and $Q$. 
Therefore, we can without loss of generality restrict the parameter space to 
the two-dimensional $(A,Q)$. 

\medskip 
Given the plethora of possibilities, it is useful again to
discuss separately the two branches with $A$ positive or negative.

%----------------------------------
\subsection{Positive branch}
%----------------------------------

Looking at Fig. \ref{Figg00A1} and the top-left panel of Fig.~\ref{FigrApos}, 
we can divide the branch with $A>0$ in two regions:
\begin{itemize}
\item Region (i), where the non-metric singularity of type~(1) is present, 
and all plots interrupt at its location $\rnm(A\beta=2)$; 
\item Region (ii), where the type~(1) singularity is absent, and 
all plots continue to $r=0$. 
\end{itemize}

\medskip 
For given $A$, the demarcation between the two regions is given by a certain 
critical value of the charge, which we call $Q_0(A)$: when below this value 
we are in region~(i), and when above we are in region~(ii). 
In region~(i), the location $\rnm$ of 
the non-metric singularity is a decreasing function of $Q$. 
This decreasing behaviour continues until the critical charge, 
e.g., $Q_0\sim 0.8$ for $A=1$, where a minimum value of $\rnm$ is reached and 
we pass in region~(ii). In region~(ii), the type~(1) singularity is replaced 
by the type~(2), and its location is now an increasing function of $Q$. 
Numerical analysis shows that the value $Q_0(A)$ which marks the turning point 
from region (i) to region (ii) is a slowly growing function of $A$. 
For instance $Q_0(0.1)\sim 0.55$, $Q_0(1)\sim 0.8$ (see the top-left panel of 
Fig.~\ref{FigrApos}) and $Q_0(12)\sim 1.2$ (the top-right panel). 

\medskip 
Next, let us move our attention to the horizons, drawn in blue. These are
zeroes of $f_*^2$ which are simultaneously infinities of $g_*^2$.
At small $Q$, the location of the event horizon is a decreasing function
of $Q$ similar to the one of general relativity, $r_+ = M + \sqrt{M^2-Q^2}$, 
which is also shown in the plot for comparison. The difference is that at each 
$Q$ the location of the event horizon is at a larger radius than in 
the case of general relativity. 
This generalises to the charged case what already seen for $Q=0$ in the 
previous section. A characteristic feature of the Reissner-Nordstr\o m 
solution is the presence of an inner horizon. 
The top-left panel of Figure \ref{FigrApos} shows that in the modified theory 
there is only a small range of parameters in which the inner horizon is 
present, and this happens within region (ii). 
Furthermore, notice that the value of the charge at which the two horizons 
merge is slightly larger than the $Q=M$ value of the Reissner-Nordstr\o m 
solution. This result suggests the possibility of screening effects of the 
electric charge in this region of parameter space. 

\begin{figure}[h]\begin{center}
\includegraphics[width=7cm]{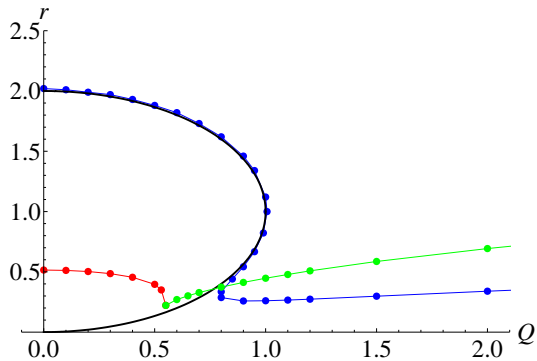} \hspace{1cm} 
\caption{\label{rA01-2}
\small{Locations of the event horizons, inner horizon, and the appearance 
of third horizon at $A=0.1$ and $M=1$. 
}}
\end{center}\end{figure}

\medskip 
A stronger departure from general relativity is the presence of a 
\emph{third} horizon:
A close look at the top-left-panel of Figure \ref{FigrApos} tells 
that for $Q$ in a narrow range around $Q=1$, we have three horizons, 
the event-, inner-, and the third-horizon
(see also right panel of Fig. \ref{Figg00A1} and Fig. \ref{fig:A3horizons} 
for the global structure of the spacetime). 
Then, when $Q$ increases, the first two (event and inner) horizons merge 
and cease to exist (as in the Reissner-Nordst\o m case), and only 
the third one continues to exist for larger value of $Q$---passed 
the critical value at which the third horizon first appears, 
and is always behind the type~(2) non-metric singularity. 
The behaviour of horizons around $Q=1$ and the appearance of the third horizon 
can be much clearly seen for smaller values of $A$ than $A=1$. 
In Figure \ref{rA01-2}, we show plots similar to Figure \ref{FigrApos} 
but with $A=0.1$. 
The locations of both the horizon and non-metric singularity grow 
with $Q$, but the singularity grows faster. This suggests that there is 
a region of parameter space in the theory where we can arbitrary overcharge 
a black hole without ever exposing the $r=0$ singularity, i.e., 
the central singularities are always behind non-metric singularities (or 
otherwise covered by an event horizon). 

\medskip 
When we increase $A$, the value $Q_0(A)$ at which we pass from region~(i) to region~(ii) slowly 
grows. Also, the location itself of the non-metric singularities grows with 
$A$. Specifically, this grows faster than the location of the event horizon, 
a phenomenon already seen in the uncharged case that led to the existence of 
a notion of ``minimal'' black hole. The same thing happens in the charged case 
and, because the presence of a charge shrinks the radius of the event horizon, 
we can expect that at fixed $M$ the non-metric singularity of type~(1) 
is exposed at smaller values of $A$ as we increase the charge. 
This situation is depicted in the top-right panel of Fig. \ref{FigrApos}, 
where we consider the value $A=12$. This value is smaller than 
the $A\simeq 14.78$ required -- for $M=1$ -- to expose the non-metric 
singularity in the uncharged case (cf. Fig. \ref{FigrUnico}), but 
can make the type~(1) singularity naked if enough charge is present 
(e.g., $Q\sim 0.8$ for $A=12$, see Figure \ref{FigrApos}). 

\medskip 
For such a large $A$ (still the top-right panel), we also notice the complete 
disappearance of the event horizon, swallowed up by the type~(1) non-metric 
singularity. This does not exclude the possibility than an horizon is still 
present beyond the singularity, but we do not have access to that region. 
If we continue to increase $A$, the event horizon disappears 
for smaller and smaller values of $Q$, until the value $A=2 e^2 M^2$ found 
in the previous Section. 
This is the value for which the event horizon disappears also in the uncharged 
case. Note however that a horizon is still present in region (ii), behind the 
non-metric singularity of type~(2). Therefore even at such large $A$, we can 
still have a horizon that covers the true singularity at the centre, 
if the charge is high enough. 

\begin{figure}[h]\begin{center}
\includegraphics[width=7cm]{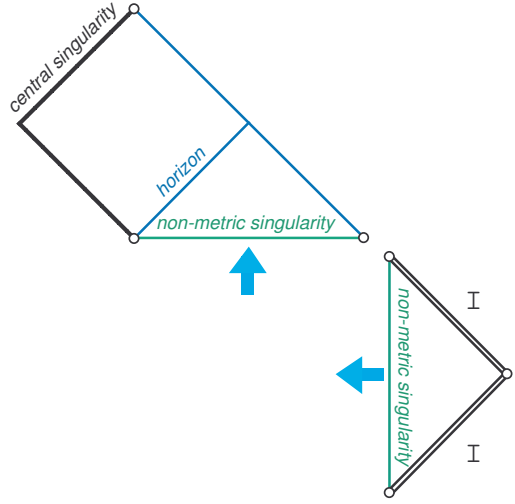} \hspace{1cm} 
\caption{\small{
Conformal diagram for the solution with $A=1$ and 
$M=1$, $Q \gtrsim 1.1$ or $A=12$, $M=1$, $Q \gtrsim 1.2$. 
See the top two panels of Fig.~\ref{FigrApos}. 
The double line, ${\cal I}$, denotes null infinity. 
There is a naked non-metric singularity of type~(2) (green line), 
and inside it, a horizon (blue line) that covers a true singularity 
at the centre (thick dashed line).  
Here, the flow of time is taken to be vertical. As a consequence, 
in this figure, the two green lines---one vertical and the other 
horizontal---should be identified, as they denote the same non-metric 
singularity. Covered by a horizon (blue line) inside the non-metric 
singularity, the central singularity at $r=0$ is non-spacelike.    
The full extension of the spacetime is obtained by gluing together 
copies of the fundamental block depicted here, at blue solid line. 
}\label{fig:A1nakedhorizon} } 
\end{center}
\end{figure}

\begin{figure}[h]\begin{center}
\includegraphics[width=7cm]{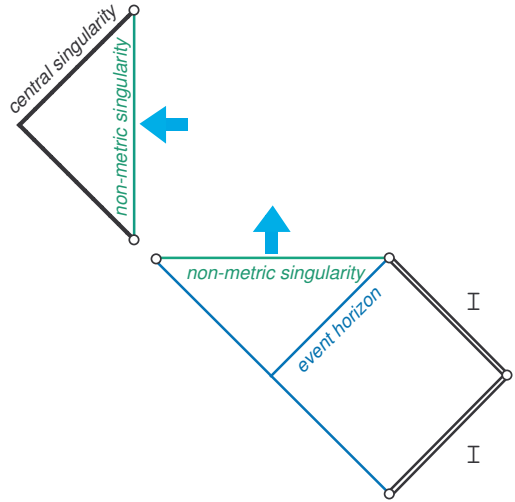} \hspace{1cm} 
\caption{\small{
Conformal diagram for the charged black hole solution with $A=1$ and 
$M=1$, $Q=0.9$. See the top-left panel of Fig.~\ref{FigrApos}. 
The double line denotes null infinity. 
There is a type~(2) non-metric singularity (green line) hidden by 
the event horizon (blue line). Inside the non-metric singularity is 
the central singularity (thick dashed line). 
The flow of time is vertical, and the two green lines---one vertical and 
the other horizontal---are identified as indicated. The central singularity 
inside the non-metric singularity is thus non-spacelike. 
The full extension of the spacetime is obtained by gluing together 
copies of the fundamental block depicted here, at blue solid line. 
}\label{fig:A1horizon} } 
\end{center}
\end{figure}

\begin{figure}[h]\begin{center}
\includegraphics[width=12cm]{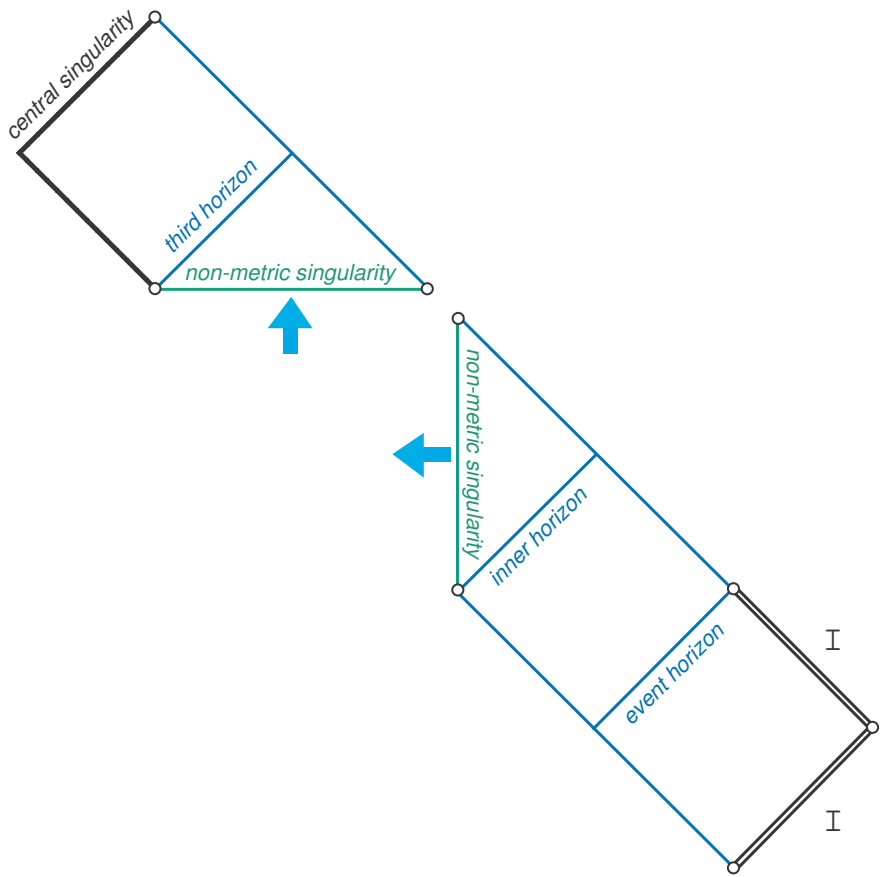} \hspace{1cm} 
\caption{\small{
Conformal diagram for the charged black hole solution with 
$A=0.1$ and $M=1$, $Q=0.9$. See Fig.~\ref{rA01-2}.
The double line denotes null infinity. 
There is a type~(2) non-metric singularity (green line) 
inside the inner horizon, and the third-horizon is located 
inside the non-metric singularity. 
The flow of time is vertical, and the two green lines---one vertical 
and the other horizontal---denoting the same non-metric 
singularity are identified. The central singularity (thick dashed line) 
inside the third-horizon is non-spacelike.  
The full extension of the spacetime is obtained by gluing together 
copies of the fundamental block depicted here, at blue solid line. 
}\label{fig:A3horizons} }  
\end{center}
\end{figure}

In Figures  \ref{fig:A1nakedhorizon}, 
\ref{fig:A1horizon} and \ref{fig:A3horizons} we report the Penrose diagrams 
of different configurations for which the solution is complete, 
i.e. $r\in(0,\infty)$. 

%-----------------------------------------------------
%%%%%%

%-----------------------------------------------------
%%%%%%

%-----------------------------------------------------
%%%%%%

%%%%%%%%%
%-----------------------------------------------------

%%%%%%%%%
%-----------------------------------------------------

\cleardoublepage

\newpage

%----------------------------------
\subsection{Negative branch}
%----------------------------------

In the negative branch the standard two horizons of the Reissner-Nordstr\o m 
solution are more easily distinguishable (see the bottom panels of 
Figure~\ref{FigrApos}). The effect of a negative $A$ to shrink the location of 
the event horizon seen in the uncharged case generalises to the charged case 
in a way that qualitatively maintains the standard dependence on $Q$. 
At the same time, the inner horizon is pushed to a larger radius. 
As a consequence, the charge at which the two horizons merge and disappear is 
smaller than the $Q=M$ value of the general relativity case. 
This hints at possible effects of antiscreening of the electric charge 
in this theory. 

\begin{figure}[h]\begin{center}
\includegraphics[width=7cm]{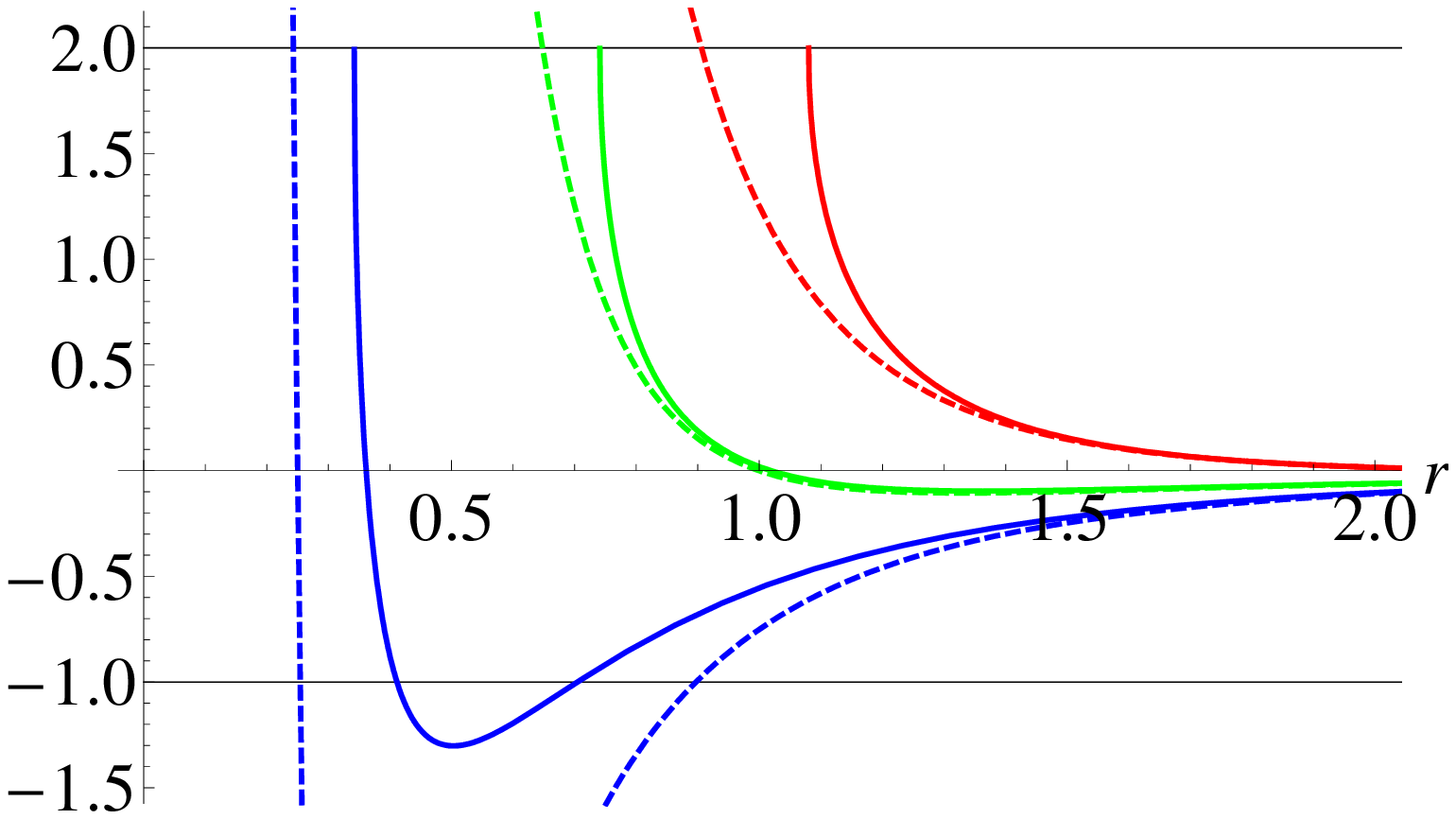} \hspace{1cm} 
\includegraphics[width=7cm]{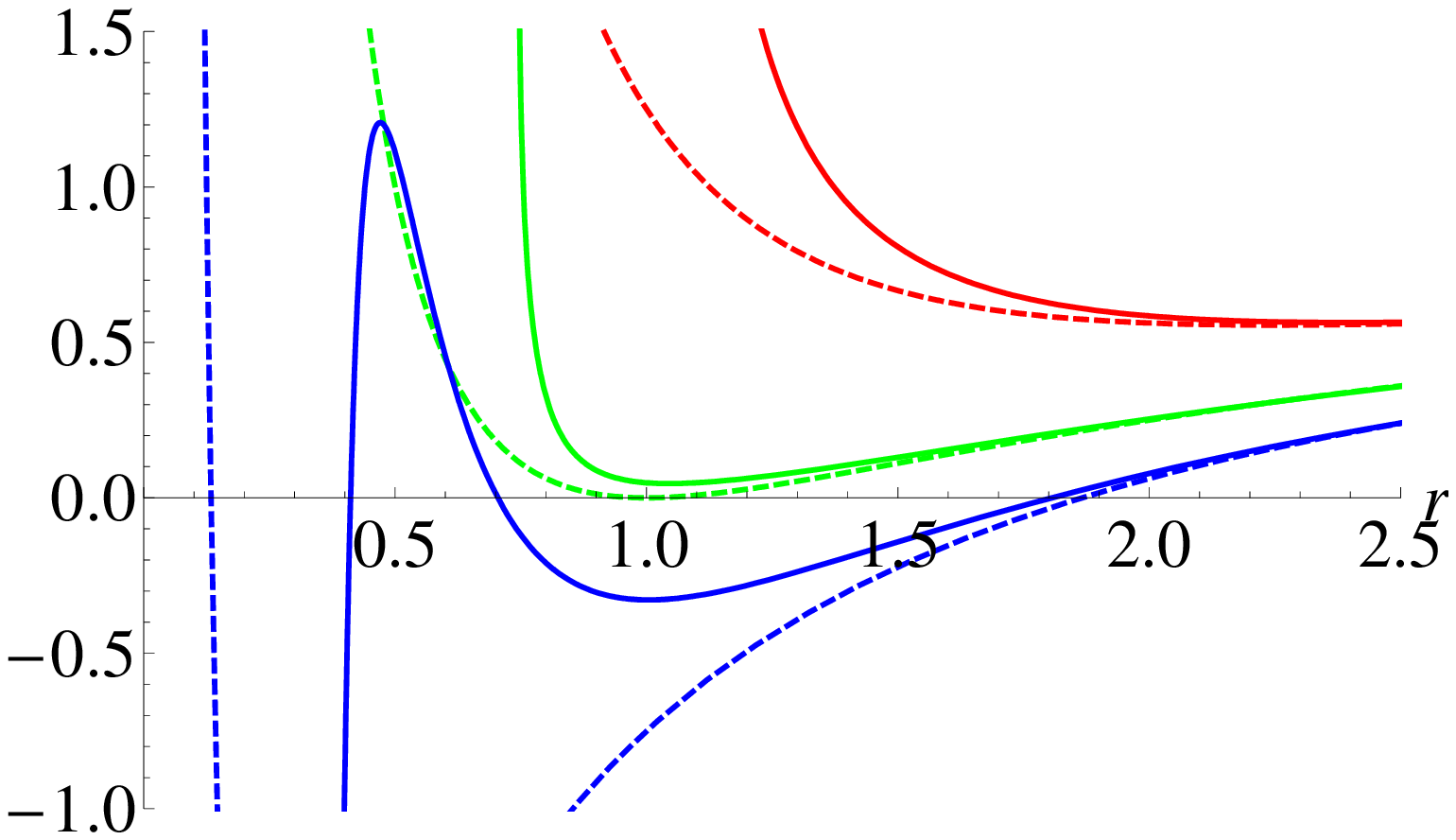}
\begin{picture}(0,0)(0,0)\put(-190,118){\large$g_{00}$}\end{picture}%\put(0,38){\large$r$}
\begin{picture}(0,0)(240,0)\put(-190,115){\large$A\beta$}\end{picture}%\put(0,62){\large$r$}
\caption{\small{
The numerical solutions for $M=1$ and $A=-1$ of the modified theory 
(solid lines) plotted against the Reissner-Nordstr\o m solution of 
general relativity (dashed lines) for $Q=0.5$ (blue), $Q=1$ (green) and 
$Q=1.5$ (red). 
Notice the matching asymptotic behaviour. $g_{00}$ always diverges at 
the radius where the $A\beta=2$ singularity occurs. 
For $Q=1$ (green line) there is no horizon, unlike in the 
Reissner-Nordstr\o m solution. For $Q=0.5$ (blue lines) the singularity at 
$A\beta=-1$ is crossed twice, and consequently the plot for $g_{00}$ shows 
two additional zeros before the divergence, which is now at minus infinity. 
}\label{Figg00AN1} }
\end{center}
\end{figure}

\medskip 
Concerning the appearance of non-metric singularities, we see from the bottom 
panels of Fig.~\ref{FigrApos} that 
\begin{itemize}
\item The solutions always possess the type~(1) singularity; 
\item The type~(2) singularities can also occur in the solutions, 
for $Q$ smaller than a certain value, $Q_1(A)$. 
\end{itemize}
Comparing the two bottom panels of Fig. \ref{FigrApos}, we see that the value 
$Q_1(A)$ at which the two singularities of type~(2) merge and disappear grows 
with $|A|$. 
The simultaneous presence of the two singularities makes the near horizon 
behaviour differ significantly from the case of general relativity: 
See the blue plots in Fig. \ref{Figg00AN1} which have $A=-1$, $M=1$ 
and $Q=0.5$. 

\medskip 
Let us look at the bottom-right panel of Fig.~\ref{FigrApos}. 
We see that when increasing $|A|$, the two horizons approach each other, 
and the extremal case happens at a smaller and smaller charge. The location of 
the outmost horizon still maintains the qualitative dependence on $Q$ of the 
Reissner-Nordstr\o m solution, but the location of the inner one has now 
a very deformed dependence. The two horizons disappear completely at the value 
$A=-54 M^2/e^2$ computed in the uncharged case, and as already seen then, 
the type (2) singularity becomes exposed before the two horizons disappear: 
the bottom-right panel shows that at $A=-7$ the horizons are still present 
but inside the singularity of type~(2).

%-----------------------------------------------------
\section{Summary}\label{SecConc} 
%-----------------------------------------------------

We have studied spherically symmetric black holes in the framework of 
Krasnov's modified self-dual gravity~\cite{Krasnov1}, in which a Weyl 
curvature dependent functional is added to the Plebansky action of general 
relativity. The remarkable novelty of this modification is that there appear 
no extra physical degrees of freedom for gravity.
This is a significant difference 
from other modified theories of gravity proposed in the literature, and 
it requires the use of a self-dual 2-form $B$ as the fundamental field: 
The metric describing the geometry of spacetime is only a derived quantity.
The derived metric can be obtained using the self-duality of $B$, and the 
field equations it satisfies are still second order
as in general relativity. At least for the case of spherical symmetry, the 
field equations can be casted in a way which is significantly similar to 
general relativity, with the effect of the modification clearly 
distinguishable. The modification increases the non-linearity of the 
equations, due to the extra feedback effect coming from the additional Weyl 
curvature dependent---otherwise ``cosmological constant''---term.

\medskip   
The first result of the paper is to show how to couple electromagnetism 
to the theory. This uses the coupling to Pleba\'{n}ski's theory developed in 
\cite{Capo2}, which is shown to apply straightforwardly also to the modified 
theory, as anticipated by Krasnov \cite{Krasnov1}. We have then proved the 
staticity of the electro-vacuum spherically symmetric solutions for any form 
of $\Phi$ (Birkhoff's theorem, Section \ref{sec:Birkhoff}).  
Under the assumption of spherical symmetry and accordingly with the staticity 
result, we have considered spherically symmetric, static black hole 
spacetimes for the two simplest profiles of $\Phi$, i.e. the linear 
modification $\Phi=-\Lambda/3 + a \beta(r)$ and the quadratic modification 
$\Phi=-\Lambda/3 + A \beta^2(r) / 2$, with $a$ and $A$ parameters and 
$\beta(r)$ the `magnitude' of Weyl curvature of our spherically symmetric 
spacetimes. In both cases, the structures near the horizon and near 
the central singularity significantly differs from general relativity, 
and the notion of extremal black hole changes with the parameters.
We summarise our main results below.

\medskip 
In the linearly modified case, the solutions do not approach asymptotically 
flat geometries at large distances even when the cosmological constant 
$\Phi_0= - \Lambda/3$ is set to zero, as we have briefly discussed in 
Section~\ref{sect:AF}, and hence the solutions have no direct physical 
relevance. 
Since the linear modification fails to satisfy a certain notion of 
analyticity, the question arises if there is a relation between the latter 
and the possibility of asymptotic flatness. Such a relation would be useful
to restrict the class of $\Phi$'s with physically relevant solutions.
The linear case is nonetheless worth examining since it can be easily solved 
analytically. The solutions obtained in Section~\ref{sec:linear} explicitly 
show how the $Q$-dependence of the locations of the horizons is affected 
by the parameter $a$. 

\medskip 
In the quadratically modified case, the non-linearity of the equations 
of motion is more involved, and we have not been able to find analytic 
solutions for $Q \neq 0$. We resorted to numerical computations to study 
the charged black holes. For this purpose, we first considered a perturbative 
approach to the field equations. The approximate solutions found in this way 
are valid at large distances, and allowed us to show that the solutions 
have the standard Reissner-Nordstr\o m asymptotic behaviour, and thus 
the parameters $(M,Q)$ appeared there can be interpreted in the standard 
sense, the ADM mass and the total charge, respectively, in accordance with 
the arguments of Section~\ref{sect:AF}.

\medskip 
The numerical methods allowed us also to further investigate the quadratic 
modification case with $Q=0=\Lambda$, whose vacuum black hole solutions had 
already been analytically found in \cite{Krasnov2}. We studied the location 
of the event horizon, and showed that a black hole of mass $M$ can only exists 
for $A\in\big(-(\f52)^3 e^{-1} M^2, 2 e^2 M^2\big)$, where the lower bound 
differs from the one proposed in \cite{Krasnov2}. 
For values of $A$ outside this interval, a non-metric singularity is naked 
outside the horizon; it is of type~(1) if $A>0$, and of type~(2) if $A<0$. 
Concerning the nature of the singularity, it was shown in \cite{Krasnov2} that 
a change of coordinates can be found in which the fundamental fields are 
finite. Therefore this is not a physical singularity for the theory, and the 
fundamental fields can be continued beyond it. Here we have introduced also a 
scalar quantity, analogue of the Ricci scalar, and showed explicitly that it 
is finite at the non-metric singularities. 
The vacuum $Q=0=\Lambda$ case results are summarised as follows. 

\begin{enumerate}
\item The location of the horizon, $\rH(A,M)$, at fixed $M$ is pushed at 
larger/smaller values than in GR according to the sign of $A$. 

\item The internal structure becomes more intricate: the true singularity is 
surrounded by the non-metric singularity and an internal horizon. 
The non-metric singularity is of type~(1), $A\beta=2$, for $A>0$ and 
type~(2), $A\beta=-1$, for $A<0$. The new internal horizon can be seen 
explicitly for $A<0$, but for $A>0$ one has to change branch of the solution 
after the singularity to see it, or go to $\beta$ as radial coordinate, as it 
is done in \cite{Krasnov2}. 

\item For $|A|$ too large, the non-metric singularity is exposed outside the 
event horizon. Therefore in a given modified action with fixed $A$, there is 
a minimal mass that a black hole can have. Beyond this limit, distant 
observers would not see the event horizon, but a non-metric singularity first.

\end{enumerate}

\medskip 
We next summarise the charged case.  
Our main focus has been on which type of (e.g., (1) $A\beta=2$ or 
(2)$ A\beta=-1$) non-metric singularities appear and whether those 
singularities are covered by the event horizon. Given a fixed $M$, the answer 
depends upon the values of the parameters $A$ and $Q$.  

\begin{enumerate}
\item As in GR, one can find an event horizon as well as an inner horizon. 
The location $\rH(A,Q,M)$ of the event horizon for fixed $M$ and $Q$ is 
pushed at larger or smaller values than in GR according to the sign of $A$. 
For the inner horizon, we have the opposite behaviour. 
As a consequence, the extremal case is now at a larger or smaller value 
than $Q=M$ as in GR. 
This suggests that in the theory, electromagnetic charges might experience 
the screening or antiscreening effects according to the sign of $A$,   
as we have discussed in Section~\ref{SecRN}. 
The physical implication of the screening effects, however, remains an open 
issue. 

\item The true singularity at $r=0$ is surrounded by non-metric singularities, 
and (if exists) inner horizons. The outmost non-metric singularity is:

\begin{itemize}
\item type (1) at small $Q$, then (2) at larger $Q$ for $A>0$

\item type (2) at small $Q$, then (1) at larger $Q$ for $A<0$
\end{itemize}

\item Increasing $|A|$ too much, the non-metric singularity is exposed 
outside the event horizon. This happens earlier and earlier, as we pour a 
larger charge into the hole. Therefore in a given modified action with 
fixed $A$, there is a minimal mass that a black hole of charge $Q$ can have, 
and this value increases as we increase $Q$. 

\item A region of parameter space exists, with $A>0$, where the type~(2) 
singularity is naked, but inside it there always exists a horizon which hides 
the singularity at $r=0$ for arbitrarily large values of the charge. 

\end{enumerate}

\medskip   
Note that there could be further structure behind all singularities of 
type~(1), which we do not attempt to access numerically. This would require 
a change of coordinates, or the knowledge of the analytic solution 
as it was done for the vacuum solution in \cite{Krasnov2}.
(Compare with Section \ref{SecVacuum} for the positive branch of the uncharged 
case.)  
It would certainly be interesting to do so, changing the radial coordinate as 
in the vacuum case. 
However, given the limitations of our understanding of analytic solutions, 
we prefer to restrain from doing so, and focus instead the attention of 
the reader on the near horizon structure. 

\medskip 
Our results also bring some insight into how it is possible to 
restrict the class of physically relevant types of $\Phi$. We showed that 
spherically symmetric black holes (charged or not) are not asymptotically flat in 
some type of $\Phi$ non-analytic at the origin, and one is tempted to 
conjecture that this generalises to any non-analytic choice. 
Although charged black holes are not relevant in astrophysical context, 
the results obtained show that a deeper understanding of the modified theory, 
as well as further useful restrictions on $\Phi$ can be found by looking at 
matter coupling. 

\medskip 
Finally, we would like to recall that given $M$, there are parameter regions  
of $(A,Q)$ in which both types of non-metric singularities can be naked.  
In this case, distant observers would see the breakdown of the standard 
spacetime picture based on the metric description outside the horizon. 
Nevertheless, from the view point of this modified self-dual gravity, 
there is nothing to be afraid of, since the non-metric singularities 
are harmless for the fundamental fields, even when naked. 
Therefore, in this modified self-dual gravity, 
the notion of the cosmic censorship itself may also need be modified. 
In this regard, we should also recall that 
a non-metric singularity is specified as a ($r=const.$) hypersurface,  
across which the time and spatial (radial)-coordinates change their spacetime 
role. This, in particular, implies that as illustrated by the conformal diagram in 
Figure~\ref{fig:A1naked} for the simplest vacuum case, 
a true singularity at the centre ($r=0$) covered by a naked non-metric 
singularity is achronal (non-timelike), unless covered by any other horizons.  
We also note from Figure~\ref{fig:A1naked} that the hypersurface of 
a naked non-metric singularity plays the role of the (partial) 
Cauchy surface for the interior region, and therefore once a causal 
curve goes across a non-metric singularity and enters the interior region, 
then it has to fall into the central singularity and never comes back to 
the region outside the non-metric singularity. 
Therefore, in this sense the hypersurface of a non-metric singularity may be 
viewed as a one-way membrane, or a type of ``horizon.'' 
This view point may appear to be in favour of the cosmic censorship 
in the sense that the true singularities of the theory appear to always be 
covered by either the event horizon, or a non-metric singular {\em horizon}. 
Note that in the present context, one cannot tell which horizon, 
future or past, a non-metric singularity corresponds to, due to the 
time-symmetric nature of static spacetimes. However, the situation would be 
clear when dynamical (non-stationary) cases are considered. 
It would be interesting to consider a sensible definition of a Cauchy 
problem for the evolution in such a spacetime and the notion of cosmic 
censorship, within the framework of the modified self-dual gravity theory.

%%%%%%%%%%%%% 
\subsection*{Acknowledgements}
%%%%%%%%%%%%%
We are thankful to Kirill Krasnov for clarifications on the conformal ambiguity,
and to Costantinos Skordis for useful discussion and numerical help.
AI would like to thank Perimeter Institute for Theoretical Physics for its 
hospitality during the time this work was initiated and large part of 
the work was carried out.  
Research at Perimeter Institute for Theoretical Physics is 
supported in part by the Government of Canada through NSERC and by 
the Province of Ontario through MRI.

%%%%%%%%%%%%%%%%%

\end{document}